\documentclass[11pt]{article}

\usepackage{acl}

\usepackage{times}
\usepackage{latexsym}
\usepackage{makecell}

\usepackage[T1]{fontenc}

\usepackage[utf8]{inputenc}

\usepackage{microtype}

\usepackage{inconsolata}
\usepackage{amsmath}
\usepackage{graphicx}
\usepackage{multirow}
\usepackage{booktabs} 
\usepackage{amssymb} 
\usepackage{dsfont}  
\usepackage{bbm}         
\usepackage[export]{adjustbox}
\usepackage[table]{xcolor}
\usepackage{tabularx,colortbl}

\definecolor{heatO5}{RGB}{255,164,105}
\definecolor{heatO4}{RGB}{255,185,135}
\definecolor{heatO3}{RGB}{255,207,170}
\definecolor{heatO2}{RGB}{255,226,205}
\definecolor{heatO1}{RGB}{255,241,230}

\definecolor{heatB1}{RGB}{232,242,250}
\definecolor{heatB2}{RGB}{207,229,247}
\definecolor{heatB3}{RGB}{175,211,241}
\definecolor{heatB4}{RGB}{139,192,235}
\definecolor{heatB5}{RGB}{103,172,226}
\definecolor{sectiongray}{RGB}{245,245,245}
\definecolor{ourslight}{RGB}{239,246,253}
\definecolor{oursstrong}{RGB}{218,233,248}
\usepackage[table]{xcolor}

\definecolor{best1}{RGB}{255,158,96}
\definecolor{best2}{RGB}{255,190,145}
\definecolor{best3}{RGB}{255,220,195}

\definecolor{bad1}{RGB}{100,170,235}
\definecolor{bad2}{RGB}{175,215,245}

\usepackage[table]{xcolor}

\definecolor{panelgray}{RGB}{240,244,247}
\definecolor{combinedblue}{RGB}{235,245,252}
\definecolor{bestorange}{RGB}{255,224,195}
\usepackage[most]{tcolorbox}
\usepackage{xcolor}

\definecolor{casebg}{RGB}{248,249,250}
\definecolor{caseborder}{RGB}{205,211,218}
\definecolor{caseheader}{RGB}{239,242,245}

\definecolor{casegreen}{RGB}{232,247,238}
\definecolor{casered}{RGB}{252,235,235}
\definecolor{caseyellow}{RGB}{255,246,214}

\newtcolorbox{casebox}[1]{
    enhanced,
    colback=casebg,
    colframe=caseborder,
    boxrule=0.5pt,
    arc=1.2mm,
    left=1.5mm,
    right=1.5mm,
    top=1.2mm,
    bottom=1.2mm,
    before skip=1.5mm,
    after skip=1.5mm,
    title={#1 },
    coltitle=black,
    fonttitle=\small\bfseries,
    colbacktitle=caseheader,
    attach boxed title to top left={
        xshift=1.5mm,
        yshift=-1mm
    },
    boxed title style={
        boxrule=0pt,
        arc=1mm,
        colback=caseheader,
        left=1mm,
        right=1mm,
        top=0.4mm,
        bottom=0.4mm
    }
}

\usepackage[most]{tcolorbox}
\usepackage{tipa}
\title{X-MDD: LLM-Based Mispronunciation Detection and Diagnosis for Unseen Languages with Canonical References}

\author{
  \textbf{Xintong Wang\textsuperscript{1}},
  \textbf{Junchuan Zhao\textsuperscript{1,*}},
  \textbf{Ye Wang\textsuperscript{1,*}}
  \\[0.5em]
  \textsuperscript{1}School of Computing,
  National University of Singapore
  \\
  \{xintong.wang, junchuan\}@u.nus.edu,
  dcswangy@nus.edu.sg
}

\begin{document}
\maketitle

\begingroup
\renewcommand{\thefootnote}{\fnsymbol{footnote}}
\footnotetext[1]{Corresponding authors.}
\endgroup

\begin{abstract}
Mispronunciation detection and diagnosis (MDD) provides phone-level feedback to language learners, but extending it to new languages typically requires collecting and annotating learner speech for each target language. This repeated annotation effort is costly and limits the language coverage of pronunciation feedback. We study whether MDD capabilities learned from annotated learner speech can be applied to languages unseen during MDD training when canonical pronunciation references are available. Our multilingual LLM-based model, \textbf{X-MDD}, uses canonical speech and phones as references to predict the phone sequence actually produced by a learner. It incorporates in-context demonstrations for inference-time adaptation and is further optimized with group relative policy optimization (GRPO) using detection F1 and phone error rate. We further analyze how canonical inputs affect prediction and find that (1) canonical speech and phones jointly contribute to actual-phone prediction, and (2) conflicting speech evidence reduces the tendency to copy canonical phones. Trained only on Mandarin and English, X-MDD achieves a detection F1 of 38.07 on Arabic Qur'anic recitation, compared with 44.14 for a supervised challenge baseline trained on Arabic. It also achieves F1 scores of 78.68 and 58.78 on Mandarin and English, respectively, demonstrating effective multilingual MDD within a single model.
\end{abstract}

\section{Introduction}
Computer-assisted pronunciation training (CAPT) provides second-language (L2)
learners with automated feedback on their pronunciation
\citep{6854192,leung2019cnn,7472856,9746727}. Two common tasks in CAPT are
automatic pronunciation assessment (APA) and mispronunciation detection
and diagnosis (MDD). APA assigns pronunciation scores
\citep{10648884,10328059,10095733}, whereas MDD identifies which phones
were mispronounced and what the learner produced instead
\citep{leung2019cnn,feng2020sed,9747414,9747569}. For example, an MDD system can
indicate that the intended phone /n/ was realized as /l/. By locating
errors and identifying their actual realizations, MDD provides specific
feedback that learners can use to improve their pronunciation
\citep{ehsani1998speech}.

Extending this feedback across languages remains challenging. Existing
MDD systems typically rely on annotated learner speech from the target
language \citep{peng21e_interspeech,wang24la_interspeech,
yang2026fusion}. Consequently, extending an MDD system to a new
language often requires collecting and annotating another learner-speech
corpus. Producing these annotations is labor-intensive because
annotators must identify the phones actually pronounced, including
non-native realizations influenced by the learner's first language
\citep{liu2023leveraging}. This repeated effort limits the language coverage of pronunciation
feedback. Multilingual modeling offers a way to share supervision across languages, but supporting several training languages does not by itself establish whether the learned diagnostic capability can transfer to a language unseen during MDD training. We therefore ask whether diagnostic capabilities learned from annotated learner speech in source languages can be applied to a new language without target-language MDD training, given its canonical pronunciation references.

Recent progress in multilingual automatic speech recognition (ASR)
suggests a route to such transfer. Recent speech models can leverage audio--text demonstrations through in-context learning (ICL) to adapt to new speakers and language varieties without parameter updates \citep{wang2024can,roll2025context}. Recent
multilingual ASR work further demonstrates recognition of new languages
using a number of contextual examples
\citep{omnilingual2025omnilingual}. These findings motivate using demonstrations to transfer pronunciation diagnosis to unseen languages.

LLMs have also been applied to MDD, using learner speech and canonical
phone sequences to predict the phones actually pronounced
\citep{wu24l_interspeech,wu2025integrating}. However, these results do not yet establish whether such diagnostic capabilities can transfer to a language unseen during MDD training when its canonical pronunciation is available. Two issues require attention. First, the amount and
language coverage of annotated learner speech are much smaller than
those of multilingual ASR data, so learning to exploit demonstrations
for MDD must rely on limited task supervision. Second, canonical
information must support recognition while preserving sensitivity to
pronunciation errors. Canonical phones specify the intended sequence,
but unfamiliar phone symbols may not provide sufficient information
about their acoustic realization. At the same time, explicit canonical priors can bias predictions toward the intended pronunciation, leading to lower recall \citep{geng2026beyond}. This creates a need for both acoustic grounding
of canonical phones and training objectives that explicitly reward
accurate error detection and diagnosis.

To address these issues, we introduce \textbf{X-MDD}, a multilingual
LLM-based MDD framework that uses canonical speech and phones as references. Given learner speech, synthesized canonical
speech, and the canonical phones, the model autoregressively
predicts the learner's actual phones. Canonical speech provides an acoustic realization of the intended pronunciation. During
supervised fine-tuning, we include in-context demonstrations so that
the model learns to use examples of actual pronunciation. We further optimize
the model using group relative policy optimization (GRPO)
\citep{shao2024deepseekmath} with a reward based on detection F1 and
phone error rate (PER), aligning training with both mispronunciation
detection and accurate phone recognition.

We train \textbf{X-MDD} on Mandarin and English and evaluate its transfer to Arabic Qur'anic recitation using QuranMB.v2~\citep{kheir26b_interspeech}. With canonical references and 128 in-context demonstrations, the model achieves a detection F1 of 38.07 on Arabic, compared with 44.14 for the supervised official challenge baseline \citep{kheir26b_interspeech}. GRPO further improves MDD performance over supervised fine-tuning across all three languages.

Our contributions are threefold:
\begin{itemize}
    \item \textbf{MDD for languages unseen during training.}
    We investigate whether diagnostic capabilities learned from source languages can be applied to a new language without target-language MDD training, given its canonical pronunciation references and in-context demonstrations for adaptation.

    \item \textbf{MDD with acoustic and phonetic references.}
    We introduce \textbf{X-MDD}, a multilingual LLM-based
    framework that combines canonical speech and phones in-context prompting and GRPO
    for actual phone recognition and mispronunciation
    detection.

    \item \textbf{Demonstration of unseen-language transfer.}
    We demonstrate transfer to Arabic without Arabic MDD model training and analyze how canonical references,
    in-context prompting, and GRPO shape the model's
    diagnostic behavior.
\end{itemize}

\section{Related Work}
\subsection{Mispronunciation Detection and Diagnosis}

Early MDD systems extended speech recognition to explicitly model common L2 pronunciation errors using extended recognition networks (ERN)~\citep{DBLP:conf/slte/HarrisonLQM09,qian2010discriminative}. Later approaches incorporated multiple sources of pronunciation information, including acoustic, graphemic, and phonemic features~\citep{li2016mispronunciation}. More recent end-to-end systems directly predict phone sequences without precomputed phone-level alignments~\citep{leung2019cnn,peng21e_interspeech,DBLP:conf/interspeech/WuLLM21,wang24la_interspeech}. Canonical text and phones have also been incorporated as linguistic priors~\citep{feng2020sed,DBLP:conf/icassp/YeMSWXTW22,DBLP:conf/icassp/KorzekwaLZCDK21}, and discriminative objectives have been developed to directly optimize detection F1~\citep{yan2022maximum}. Extending MDD across languages, L1-MultiMDD~\cite{DBLP:conf/icassp/KheirC024} jointly models multiple target languages while incorporating learners' native-language backgrounds. Although it evaluates transfer to additional learner corpora, the target languages remain represented during MDD training. Our work investigates MDD for a target language excluded from MDD training, given its canonical pronunciation references.

\subsection{LLM-Based Mispronunciation Detection and Diagnosis}

Recent work has formulated MDD as phone-sequence generation with LLMs. ATP-LLM~\citep{wu24l_interspeech} combines learner speech representations with canonical phone and text sequences to predict the phones actually pronounced. PP-ATP~\citep{wu2025integrating} extends this formulation by augmenting the textual prompt with potential pronunciations, including common mispronunciations. Related studies explore multimodal LLMs for pronunciation scoring at different granularities~\citep{DBLP:journals/corr/abs-2407-09209,wang2025fine,ahn2025english}. These studies primarily evaluate supervised adaptation within individual target languages. In contrast, \textbf{X-MDD} combines canonical speech and phones with in-context demonstrations for unseen-language MDD.

\subsection{Multilingual ASR and Phone Recognition}

In-context learning offers a way to adapt speech recognition through examples supplied at inference. \citep{wang2024can} show that Whisper can use labeled speech examples to adapt to new speakers and language varieties. \citep{roll2025context} further demonstrate the value of same-speaker examples for in-context adaptation with Phi-4 Multimodal. At the language level, Omnilingual ASR~\citep{omnilingual2025omnilingual} supports adaptation to unseen languages from a set of speech--text demonstrations. These findings motivate our use of in-context prompting for pronunciation diagnosis in a new language, where canonical pronunciation references are also available.

Multilingual phone recognition has also explored transfer to unseen languages. Allophant~\citep{DBLP:conf/interspeech/GlockerHG23} and \citep{magoshi25_interspeech} use articulatory attributes to support phone recognition in unseen languages. ZIPA~\citep{DBLP:conf/acl/ZhuSCM25} and PhoneticXEUS~\citep{bharadwaj2026empirical} develop International Phonetic Alphabet (IPA)-based phone recognizers using Connectionist Temporal Classification (CTC) or transducer objectives, with evaluations covering unseen languages and pronunciation variation. Our work instead studies MDD for languages excluded from MDD training, where the predicted phones must reveal deviations from an utterance's canonical pronunciation. We retain each dataset's phone notation and encode its symbols with the LLM's existing Byte Pair Encoding (BPE) tokenizer, without explicitly converting the annotations to a shared IPA inventory.

\section{Method}
\subsection{Model Overview}
\label{subsec:model_overview}

\begin{figure*}[t]
    \centering
    \includegraphics[width=\textwidth]{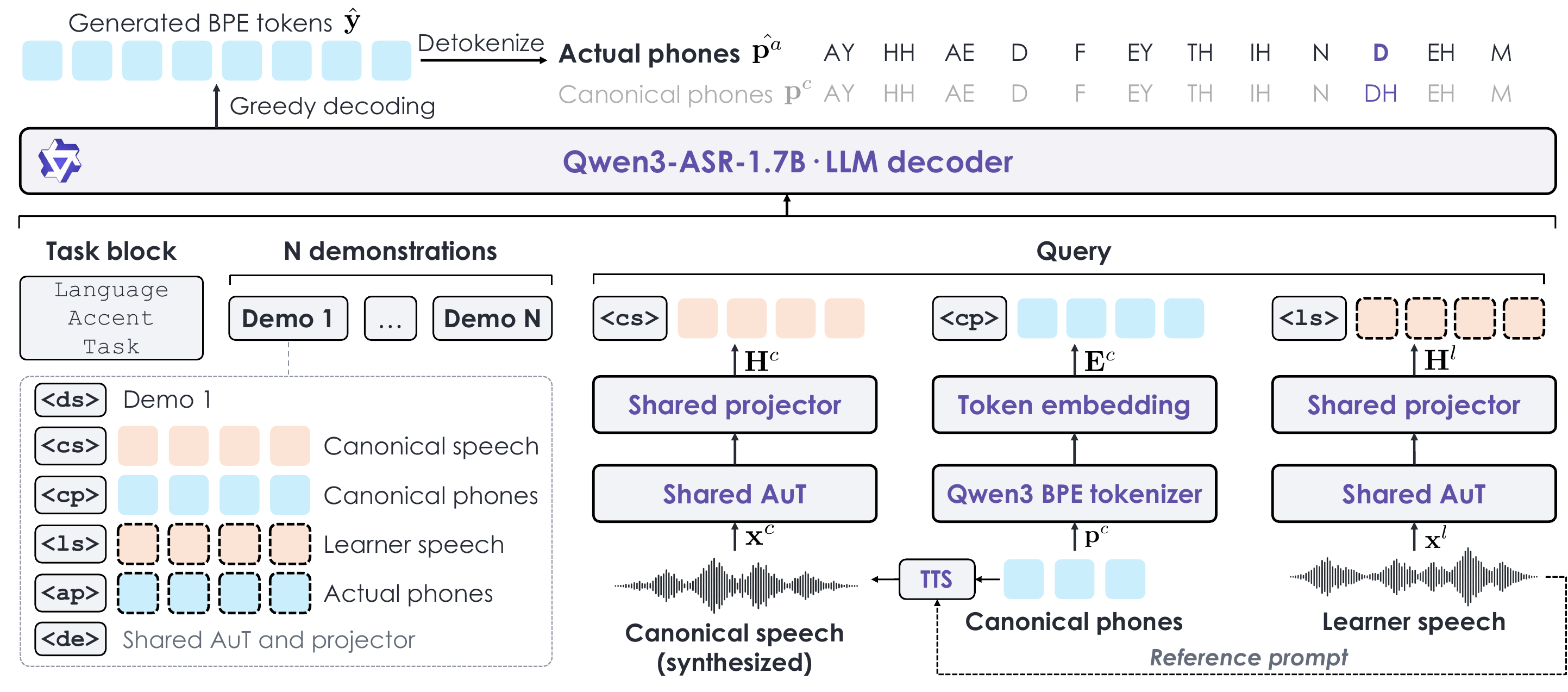}
    \caption{
        Overview of \textbf{X-MDD}, built on Qwen3-ASR-1.7B.
        Canonical speech is synthesized from canonical phones,
        using learner speech as the reference for voice cloning.
        Conditioned on these three inputs and optional in-context
        demonstrations, the model generates the learner's actual
        phone sequence for MDD.
    }
    \label{fig:model}
\end{figure*}

As shown in Figure~\ref{fig:model}, \textbf{X-MDD} comprises an
audio transformer (AuT), an audio projector, and an LLM decoder,
all initialized from Qwen3-ASR-1.7B\footnote{\url{https://huggingface.co/Qwen/Qwen3-ASR-1.7B}} \citep{shi2026qwen3}.
For each query, a text-to-speech (TTS) system synthesizes
canonical speech $\mathbf{x}^{c}$ using the canonical phone
sequence $\mathbf{p}^{c}$ and the learner's
speech $\mathbf{x}^{l}$ as the reference for voice cloning.
The model receives these three inputs together with optional
in-context demonstrations
$\mathcal{D}_{N}=(d_{1},\ldots,d_{N})$.
Each demonstration is a tuple
$d_i=(\mathbf{x}^{c}_{i},\mathbf{p}^{c}_{i},
\mathbf{x}^{l}_{i},\mathbf{p}^{a}_{i})$,
where $\mathbf{p}^{a}_{i}$ denotes the learner's actual phone
sequence.
The canonical and learner speech share the same AuT and projector, yielding
$\mathbf{H}^{v}=\mathrm{Proj}(\mathrm{AuT}(\mathbf{x}^{v}))$
for $v\in\{c,l\}$.
Phone sequences are serialized as space-separated symbols and
encoded with the backbone's BPE tokenizer. The canonical phone
embeddings are
$\mathbf{E}^{c}=\mathrm{Emb}(\mathrm{BPE}(\mathbf{p}^{c}))$.

The decoder input combines the task instructions, demonstrations,
and query inputs into a single embedding sequence:
\begin{equation}
\mathbf{E}
=
\big[
\mathbf{E}^{\mathrm{task}};
\mathbf{E}_{1:N}^{\mathrm{demo}};
\mathbf{H}^{c};
\mathbf{E}^{c};
\mathbf{H}^{l}
\big],
\label{eq:prompt}
\end{equation}
where $\mathbf{E}^{\mathrm{task}}$ encodes the task instructions
and $\mathbf{E}_{1:N}^{\mathrm{demo}}$ concatenates the
demonstration embeddings, and $\mathbf{E}_{1:N}^{\mathrm{demo}}$ is empty when $N=0$.
Each canonical speech, canonical phone, and learner speech segment is preceded by a learnable boundary token. The boundary token embeddings are omitted from Eq.~\ref{eq:prompt} for clarity.
Given $\mathbf{E}$, the decoder autoregressively generates the
BPE token sequence $\hat{\mathbf{y}}$, with the target defined
as $\mathbf{y}=\mathrm{BPE}(\mathbf{p}^{a})$, where
$\mathbf{p}^{a}$ denotes the target actual phones.
The generated tokens are detokenized into the actual predicted phone 
sequence $\hat{\mathbf{p}}^{a}$, which is aligned with
$\mathbf{p}^{c}$ to obtain the detection and diagnosis results.

\subsection{Training}
\label{subsec:training}

\paragraph{Supervised fine-tuning.}
We fine-tune all parameters of the MDD model except the three
Conv2D layers at the input of the AuT, which remain frozen. Given the target BPE token sequence
$\mathbf{y}=(y_1,\ldots,y_T)$ and the input embeddings
$\mathbf{E}$, we minimize the negative log-likelihood:
\begin{equation}
\mathcal{L}_{\mathrm{SFT}}
=
-\sum_{t=1}^{T}
\log p_{\theta}
\left(y_t\mid\mathbf{y}_{<t},\mathbf{E}\right),
\label{eq:sft}
\end{equation}
where $p_\theta$ denotes the conditional token distribution
parameterized by the MDD model. The loss is computed only over the query response.

\paragraph{Reinforcement learning.}
\label{sec:grpo}
Starting from the SFT model, we further optimize
mispronunciation detection and phone recognition using group
relative policy optimization (GRPO)~\citep{shao2024deepseekmath}. Let $\pi_\theta$ denote the current autoregressive policy parameterized
by $\theta$. We sample $G$ token sequences
$\hat{\mathbf{y}}_j=(\hat{y}_{j,1},\ldots,\hat{y}_{j,T_j})$
from a snapshot of the current policy,
$\pi_{\theta_{\mathrm{old}}}$, using ancestral sampling with
temperature $\tau=1$.
Each sample is detokenized for reward computation.
The reward combines utterance-level detection F1 and phone
error rate (PER):
\begin{equation}
r_j
=
\lambda\,\mathrm{F1}_j
+
(1-\lambda)\max\left(0,1-\mathrm{PER}_j\right),
\label{eq:reward}
\end{equation}
where $\lambda$ balances the two reward terms.
We normalize the rewards within each group to obtain the
advantage $A_j=(r_j-\mu_r)/(\sigma_r+\epsilon)$, where
$\mu_r$ and $\sigma_r$ are the mean and standard deviation
of the group rewards, and $\epsilon$ is a small positive
constant for numerical stability.
The advantage $A_j$ is shared by all tokens of sample $j$.

We update the policy using a length-normalized surrogate
loss with a Kullback–Leibler (KL) penalty toward the frozen SFT reference
model $\pi_{\mathrm{ref}}$:
\begin{multline}
\mathcal{L}_{\mathrm{GRPO}}
=
-\frac{1}{G}
\sum_{j=1}^{G}
\frac{1}{T_j}
\sum_{t=1}^{T_j}
\\
\times
\left[
\frac{
\pi_{\theta}
\left(\hat{y}_{j,t}\mid
\hat{\mathbf{y}}_{j,<t},\mathbf{E}\right)
}{
\pi_{\theta_{\mathrm{old}}}
\left(\hat{y}_{j,t}\mid
\hat{\mathbf{y}}_{j,<t},\mathbf{E}\right)
}
A_j
-
\beta\,\mathbb{D}_{\mathrm{KL}}^{j,t}
\right],
\label{eq:grpo}
\end{multline}
where $\beta$ controls the strength of KL regularization.
The per-token KL estimator is defined as: 
\begin{equation}
\begin{aligned}
\mathbb{D}_{\mathrm{KL}}^{j,t}
={}&
\frac{
\pi_{\mathrm{ref}}
\left(\hat{y}_{j,t}\mid
\hat{\mathbf{y}}_{j,<t},\mathbf{E}\right)
}{
\pi_{\theta}
\left(\hat{y}_{j,t}\mid
\hat{\mathbf{y}}_{j,<t},\mathbf{E}\right)
}
\\
&-
\log
\frac{
\pi_{\mathrm{ref}}
\left(\hat{y}_{j,t}\mid
\hat{\mathbf{y}}_{j,<t},\mathbf{E}\right)
}{
\pi_{\theta}
\left(\hat{y}_{j,t}\mid
\hat{\mathbf{y}}_{j,<t},\mathbf{E}\right)
}
-1.
\end{aligned}
\label{eq:kl}
\end{equation}

Each sampled group is used for a single policy update.
The current and sampling policies coincide at gradient
evaluation, making clipping inactive;
Eq.~\ref{eq:grpo} therefore omits the minimum and clipping
operations, following~\citep{shao2024deepseekmath}.
The sampling-policy probabilities and advantages are
treated as constants during differentiation, while the
current policy probabilities remain differentiable with respect
to $\theta$.

\subsection{In-Context Learning via Prompting}
\label{subsec:icl}

In-context prompting allows \textsc{X-MDD} to condition its
predictions on annotated examples.
During training, each sample consists of a query containing
canonical speech, canonical phones, and learner speech, together
with $N$ demonstrations $\mathcal{D}_N$ sampled from other
utterances of the same-speaker.

At inference, all test speakers are unseen during training.
For Mandarin, we construct a support set $\mathcal{S}$ of 150
annotated utterances from the corresponding test speaker.
For English and Arabic, where same-speaker support data are
unavailable, we follow~\citep{wang2024can} and
use $k$-nearest neighbors (KNN) to retrieve demonstrations from
native-speaker support sets: SpeechOcean762~\citep{zhang2021speechocean762} for English and
\texttt{Iqra\_Extra\_IS26}\footnote{\url{https://huggingface.co/datasets/IqraEval/Iqra_Extra_IS26}}
for Arabic. The \texttt{Iqra\_Extra\_IS26} is used only at inference and
not for model training.

The prompt follows the dialogue format of Qwen3-ASR.
The task block, demonstrations $\mathcal{D}_N$, and query are included in a single user turn.
The task block specifies the target language, the query
speaker's accent, and the task description.
Each demonstration contains canonical speech, canonical phones,
and learner speech, followed by the annotated actual phones
after \texttt{<|actual\_phone|>}.
Demonstrations are enclosed by \texttt{<|demo\_start|>} and
\texttt{<|demo\_end|>}. These boundary tokens have learnable
embeddings.
All demonstration speech is encoded by the same AuT and
projector used for the query.
The query contains only the three inputs, with its actual
phones predicted in the subsequent assistant turn.
The assistant response begins with the Qwen3-ASR language
prefix, after which the decoder generates
$\hat{\mathbf{y}}$ autoregressively using greedy decoding
until \texttt{<|im\_end|>} is emitted.
The generated tokens are detokenized to obtain
$\hat{\mathbf{p}}^{a}$. Further details are provided in Appendix~\ref{sec:details_app}.

\section{Experimental Setup}
\subsection{Datasets}
\label{sec:data}


\paragraph{Training data.}
For Mandarin, we combine native speech from
AISHELL-3~\citep{shi2020aishell} with human-annotated
Sichuan-, Singaporean- from
MagicHub\footnote{\url{https://magichub.com/datasets/}}
and LATIC~\citep{mqtj-qh10-21}.
For English, we combine native speech from
TIMIT~\citep{garofolo1993timit} with non-native speech from
SpeechOcean762~\citep{zhang2021speechocean762}
(Mandarin L1) and
L2-ARCTIC~\citep{DBLP:conf/interspeech/ZhaoSSLCLG18}
(six L1 backgrounds). Further details on the datasets are provided in
Appendix~\ref{sec:datasets_app}. We use Pinyin initials and tonal finals for Mandarin
and the stress-free 39-phone CMU inventory for English. 

\paragraph{Evaluation data.}
The Mandarin evaluation set contains 900 utterances from
six speakers: four with accents represented in training
(Sichuan, Singaporean, and Russian) and two with held-out
accents (Korean and Chadian).
Following~\citep{peng21e_interspeech}, we hold out one
L2-ARCTIC speaker per L1 background for English evaluation,
yielding 900 utterances from 6 speakers.
For Arabic, we use the official QuranMB.v2 test set from
the IqraEval 2026 challenge. Further details on the datasets are provided in
Appendix~\ref{sec:datasets_app}.

\subsection{Implementation Details}
\paragraph{Canonical speech synthesis.}
For Mandarin and English, we synthesize canonical speech
using CosyVoice3~\citep{du2025cosyvoice}
(Fun-CosyVoice3-0.5B)\footnote{
\url{https://huggingface.co/FunAudioLLM/Fun-CosyVoice3-0.5B-2512}}, 
with the corresponding learner recording as the voice-cloning
reference.
For Arabic, we use SILMA TTS\footnote{
\url{https://huggingface.co/silma-ai/silma-tts}},
a pretrained Arabic--English TTS model.

\paragraph{Training and inference.}
We initialize the MDD backbone from Qwen3-ASR-1.7B and train
on two NVIDIA L40S GPUs.
SFT uses AdamW~\citep{DBLP:conf/iclr/LoshchilovH19} with
learning rates of $2\times10^{-5}$ for the LLM and projector
and $5\times10^{-6}$ for the AuT.
We train for 7,000 steps on mixed Mandarin and English data
with uniform sampling and an effective batch size of 48.
The AuT is frozen for the first 1,000 steps, while its input
Conv2D layers remain frozen throughout training.
The SFT checkpoint with the highest mean Mandarin/English
development MDD F1 initializes GRPO, which runs for 300 steps
on utterances with annotated errors.
GRPO uses $G=8$ hypotheses per prompt, a learning rate of
$1\times10^{-5}$, $\beta=0.01$, $\lambda=0.5$, and $\epsilon=10^{-4}$.
SFT uniformly samples the number of demonstrations from 0 to 4,
while GRPO uniformly samples it from 1 to 2. The inference evaluates settings with
up to 128 demonstrations. Arabic decoding is restricted to the IqraEval phone symbols occurring in the QuranMB test set, with further details provided
in Appendix~\ref{sec:details_app}.

\begin{table*}[!t]
\centering
\small
\setlength{\tabcolsep}{4pt}
\renewcommand{\arraystretch}{1.04}
\setlength{\aboverulesep}{1.5pt}
\setlength{\belowrulesep}{1.5pt}
\caption{MDD performance on Mandarin (ZH) and English (EN).
All values are percentages; P and R denote detection precision and recall.
For PER, F1, and DER, the best and second-best results across all models
are shown in \textbf{bold} and \underline{underline}, respectively.}
\label{tab:baselines}
\begin{tabularx}{\textwidth}{X*{5}{r}>{\hspace{4pt}}r*{4}{r}}
\toprule
\multirow{2}{*}[-1pt]{\textbf{Model}}
& \multicolumn{5}{c}{\textbf{Mandarin (ZH)}}
& \multicolumn{5}{c}{\textbf{English (EN)}} \\
\cmidrule(lr){2-6}\cmidrule(lr){7-11}
& PER$\downarrow$ & P$\uparrow$ & R$\uparrow$ & F1$\uparrow$ & DER$\downarrow$
& PER$\downarrow$ & P$\uparrow$ & R$\uparrow$ & F1$\uparrow$ & DER$\downarrow$ \\
\midrule
\multicolumn{11}{l}{\textit{Monolingual baselines}} \\
w2v-CTC~\citep{peng21e_interspeech}
& 23.04 & 74.17 & 78.58 & \underline{76.31} & 40.39
& 15.97 & 55.19 & 48.15 & 51.43 & 27.82 \\
Pitch-aware RNN-T~\citep{wang24la_interspeech}
& 28.08 & 68.33 & 37.98 & 48.83 & 49.20
& 16.56 & 51.63 & 49.25 & 50.41 & 29.98 \\
ATP-LLM~\citep{wu24l_interspeech}
& 20.53 & 82.37 & 69.23 & 75.23 & 35.36
& 14.20 & 63.67 & 20.42 & 30.92 & 21.23 \\
WHU-IASP~\citep{yang2026fusion}
& 23.00 & 73.51 & 78.59 & 75.97 & 39.21
& 14.96 & 55.42 & 52.75 & 54.05 & 27.85 \\
\midrule
\multicolumn{11}{l}{\textit{Multilingual baseline}} \\
L1-MultiMDD~\citep{DBLP:conf/icassp/KheirC024}
& 28.38 & 67.59 & 36.89 & 47.73 & 56.45
& 20.02 & 41.57 & 49.76 & 45.30 & 38.70 \\
\midrule
\multicolumn{11}{l}{\textit{Our models}} \\
\rowcolor{ourslight}
\textsc{X-MDD} (monolingual)
& \underline{18.65} & 83.15 & 63.97 & 72.31 & \textbf{28.65}
& \underline{13.29} & 58.85 & 53.25 & \underline{55.91} & \underline{20.50} \\
\rowcolor{oursstrong}
\textbf{\textsc{X-MDD} (multilingual)}
& \textbf{17.79} & 82.24 & 75.41 & \textbf{78.68} & \underline{31.55}
& \textbf{12.87} & 60.04 & 57.57 & \textbf{58.78} & \textbf{19.71} \\
\bottomrule
\end{tabularx}
\end{table*}

\subsection{Evaluation Metrics}
We evaluate phone recognition using phone error rate (PER),
mispronunciation detection and diagnoise using precision, recall, F1, and diagnostic error rate (DER).
DER measures the proportion of correctly detected
mispronunciations that are incorrectly diagnosed.
Higher precision, recall, and F1 indicate better performance,
as do lower PER and DER. 
Detailed definitions and formulas are provided in 
Appendix~\ref{sec:metrics_app}.

\section{Experimental Results}
\subsection{Comparison with Baselines}
\label{subsec:baseline_results}

Table~\ref{tab:baselines} compares \textsc{X-MDD} with existing
monolingual and multilingual MDD systems on Mandarin and English.
Our monolingual variant improves PER and DER over the monolingual
baselines on both languages.
Multilingual \textsc{X-MDD} achieves the best PER and F1 on both
languages and the best DER on English.
Relative to monolingual training, it improves F1 by 6.37 and
2.87 percentage points on Mandarin and English, respectively,
while reducing PER.

\subsection{Ablation Study on Seen Languages}
\label{subsec:seen_results}
\begin{table}[!t]
\centering
\footnotesize
\setlength{\tabcolsep}{2pt}
\renewcommand{\arraystretch}{1.08}
\setlength{\aboverulesep}{1.5pt}
\setlength{\belowrulesep}{1.5pt}
\caption{Effects of canonical references, ICL, and GRPO on seen languages (\%).
Base excludes ICL and GRPO; ICL uses four test-time demonstrations.
\textbf{Bold}/\underline{underline} mark the best/second-best F1
within each block and language.}
\label{tab:seen}
\begin{tabular*}{\columnwidth}{@{\extracolsep{\fill}}l*{6}{r}@{}}
\toprule
\multirow{2}{*}[-1pt]{\textbf{Setting}}
& \multicolumn{3}{c}{\textbf{ZH}}
& \multicolumn{3}{c}{\textbf{EN}} \\
\cmidrule(lr){2-4}\cmidrule(lr){5-7}
& PER$\downarrow$ & F1$\uparrow$ & DER$\downarrow$
& PER$\downarrow$ & F1$\uparrow$ & DER$\downarrow$ \\
\midrule
\multicolumn{7}{l}{\textit{No canonical reference}} \\
Base
& 17.75 & 79.49 & 31.67 & 15.64 & 54.89 & 25.39 \\
\quad + ICL
& 17.86 & \underline{80.21} & 32.38 & 15.39 & \underline{55.10} & 24.93 \\
\quad + GRPO
& 18.53 & 79.42 & 32.94 & 16.69 & 54.14 & 25.62 \\
\quad + ICL + GRPO
& 18.32 & \textbf{80.35} & 32.98 & 16.60 & \textbf{55.82} & 24.26 \\
\midrule
\multicolumn{7}{l}{\textit{Canonical phones}} \\
Base
& 17.95 & \underline{74.67} & 30.65 & 12.02 & 53.35 & 18.10 \\
\quad + ICL
& 20.01 & 61.75 & 25.51 & 11.90 & 54.00 & 18.05 \\
\quad + GRPO
& 18.58 & \textbf{78.03} & 33.56 & 14.27 & \underline{56.49} & 20.60 \\
\quad + ICL + GRPO
& 18.94 & 74.64 & 31.52 & 12.89 & \textbf{56.53} & 20.15 \\
\midrule
\multicolumn{7}{l}{\textit{Canonical speech}} \\
Base
& 17.17 & 77.28 & 29.66 & 14.71 & 54.29 & 21.62 \\
\quad + ICL
& 17.63 & 76.20 & 30.75 & 14.51 & 54.06 & 22.41 \\
\quad + GRPO
& 17.99 & \textbf{78.26} & 32.18 & 16.36 & \underline{54.78} & 23.14 \\
\quad + ICL + GRPO
& 18.12 & \underline{77.29} & 31.61 & 15.80 & \textbf{55.18} & 23.80 \\
\midrule
\multicolumn{7}{l}{\textit{Canonical speech + phones}} \\
Base
& 18.21 & 73.91 & 30.45 & 11.99 & 54.47 & 17.83 \\
\quad + ICL
& 17.68 & 75.87 & 30.24 & 11.64 & 54.75 & 16.86 \\
\quad + GRPO
& 18.99 & \textbf{79.34} & 34.16 & 12.91 & \underline{58.25} & 20.36 \\
\quad + ICL + GRPO
& 17.66 & \underline{79.00} & 31.32 & 12.86 & \textbf{58.88} & 19.47 \\
\midrule
\multicolumn{7}{l}{\textit{Canonical speech + text}} \\
Base
& 17.69 & 76.35 & 30.83 & 14.06 & 53.99 & 21.39 \\
\quad + ICL
& 17.39 & 77.35 & 30.74 & 13.97 & 53.92 & 22.16 \\
\quad + GRPO
& 17.74 & \underline{78.85} & 31.62 & 15.20 & \textbf{56.75} & 21.78 \\
\quad + ICL + GRPO
& 17.89 & \textbf{79.12} & 32.35 & 15.43 & \underline{55.53} & 22.46 \\
\bottomrule
\end{tabular*}
\end{table}

Table~\ref{tab:seen} examines the effects of canonical references,
ICL, and GRPO on the seen languages.
Canonical references alone (base) do not consistently improve detection:
canonical phones reduce F1 on both ZH and EN, while canonical
speech yields comparable performance to the no-reference setting.
In contrast, GRPO consistently improves F1 across all canonical reference
settings, with particularly large gains when canonical speech and
phones are combined.
ICL provides smaller and less consistent improvements, but further
improves GRPO in several settings.
For EN, combining canonical speech and phones with ICL and GRPO
achieves the highest F1, while also maintaining relatively
low PER and DER.
For ZH, the no-reference model with ICL and GRPO achieves the highest
F1 of 80.35, although canonical speech and phones with GRPO reaches
a comparable 79.34.

\subsection{Transfer to an Unseen Language}
\label{subsec:unseen_results}
\begin{table}[!t]
\centering
\footnotesize
\setlength{\tabcolsep}{4pt}
\renewcommand{\arraystretch}{1.08}
\setlength{\aboverulesep}{1.5pt}
\setlength{\belowrulesep}{1.5pt}
\caption{
Results on unseen Arabic (\%).
Base excludes ICL and GRPO; ICL uses 128 test-time demonstrations.
\textbf{Bold}/\underline{underline} mark the best/second-best
results for each metric, excluding the Arabic-trained baseline ($^\ast$).
}
\label{tab:unseen_models}
\begin{tabular}{@{}lrrr@{}}
\toprule
\textbf{Setting}
& PER$\downarrow$ & F1$\uparrow$ & DER$\downarrow$ \\
\midrule

IqraEval baseline$^\ast$
& 16.15 & 44.14 & 38.8 \\
\midrule

\multicolumn{4}{l}{\textit{No canonical reference}} \\
Base
& 606.10 & 12.72 & 92.51 \\
\quad + ICL + GRPO
& 84.64 & 18.80 & 58.93 \\
\midrule

\multicolumn{4}{l}{\textit{Canonical phones}} \\
Base
& 10.46 & 17.40 & 46.11 \\
\quad + ICL + GRPO
& 10.47 & \underline{25.06} & \underline{41.58} \\
\midrule

\multicolumn{4}{l}{\textit{Canonical speech}} \\
Base
& 462.54 & 12.55 & 93.08 \\
\quad + ICL + GRPO
& 94.38 & 17.82 & 66.80 \\
\midrule

\multicolumn{4}{l}{\textit{Canonical speech + phones}} \\
Base
& \underline{10.15} & 23.88 & 50.98 \\
\quad + ICL + GRPO
& \textbf{7.75} & \textbf{38.07} & \textbf{40.33} \\
\midrule

\multicolumn{4}{l}{\textit{Canonical speech + text}} \\
Base
& 409.68 & 13.47 & 88.04 \\
\quad + ICL + GRPO
& 95.49 & 18.58 & 66.55 \\
\bottomrule
\end{tabular}
\end{table}
\begin{table}[!t]
\centering
\footnotesize
\setlength{\tabcolsep}{4pt}
\renewcommand{\arraystretch}{1.08}
\setlength{\aboverulesep}{1.5pt}
\setlength{\belowrulesep}{1.5pt}
\caption{
Effect of the number of ICL demonstrations ($N$) on Arabic MDD.
Metrics are percentages; P/R denote precision/recall.
\textbf{Bold}/\underline{underline} mark the best/second-best
results, including ties.
}
\label{tab:unseen_icl}
\begin{tabular}{@{}rrrrrr@{}}
\toprule
$N$
& PER$\downarrow$ & P$\uparrow$ & R$\uparrow$
& F1$\uparrow$ & DER$\downarrow$ \\
\midrule
0
& 10.29 & 38.39 & \textbf{32.54}
& 35.22 & 44.55 \\
2
& 9.43 & 36.30 & \underline{32.15}
& 34.10 & 43.51 \\
8
& 8.45 & 40.85 & \textbf{32.54}
& 36.22 & 42.53 \\
32
& 8.14 & 45.92 & 31.52
& 37.38 & \underline{38.88} \\
64
& \textbf{7.38} & \underline{49.68} & 30.09
& \underline{37.48} & \textbf{37.86} \\
128
& \underline{7.75} & \textbf{49.74} & 30.84
& \textbf{38.07} & 40.33 \\
\bottomrule
\end{tabular}
\end{table}

Table~\ref{tab:unseen_models} evaluates transfer to Arabic,
which is excluded from MDD training. Canonical phones are
critical in this setting: models without them exhibit very
high PER, even with ICL and GRPO. Combining canonical speech
and phones yields the strongest results, with ICL and GRPO
further improving all three metrics. With 128 inference-time demonstrations, this configuration
achieves 38.07 F1, compared with 44.14 for the challenge
baseline trained on Arabic, showing that diagnostic capabilities
learned from Mandarin and English can transfer to Arabic when
canonical references and labeled demonstrations are available.
Table~\ref{tab:unseen_icl} further examines demonstration count.
More demonstrations generally improve F1 through higher
precision, accompanied by slightly lower recall, although
the gains are not monotonic. F1 peaks with 128 demonstrations,
whereas PER and DER are lowest with 64. A qualitative example of this trend is provided in Case~2 of
Appendix~\ref{sec:case_studies}.

\subsection{Reliance on Canonical References}
\label{subsec:canonical_analysis}
\begin{figure}[!t]
\centering
\includegraphics[width=\columnwidth]{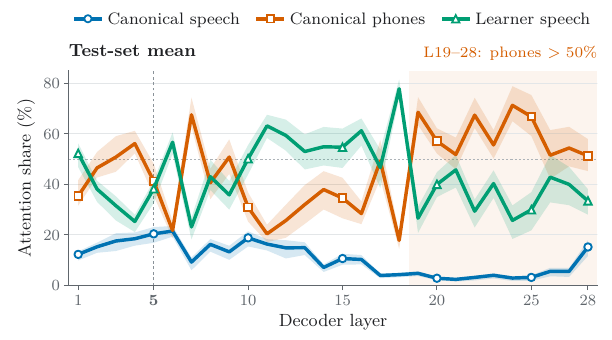}
\caption{
Mean attention shares across the three inputs on the Arabic
test set. Shading marks layers 19--28, where canonical phones
receive over 50\%. The vertical guide marks layer 5
(Figure~\ref{fig:attention_maps}).
}
\label{fig:attention_share}
\end{figure}

\begin{figure}[!t]
\centering
\includegraphics[width=\columnwidth]{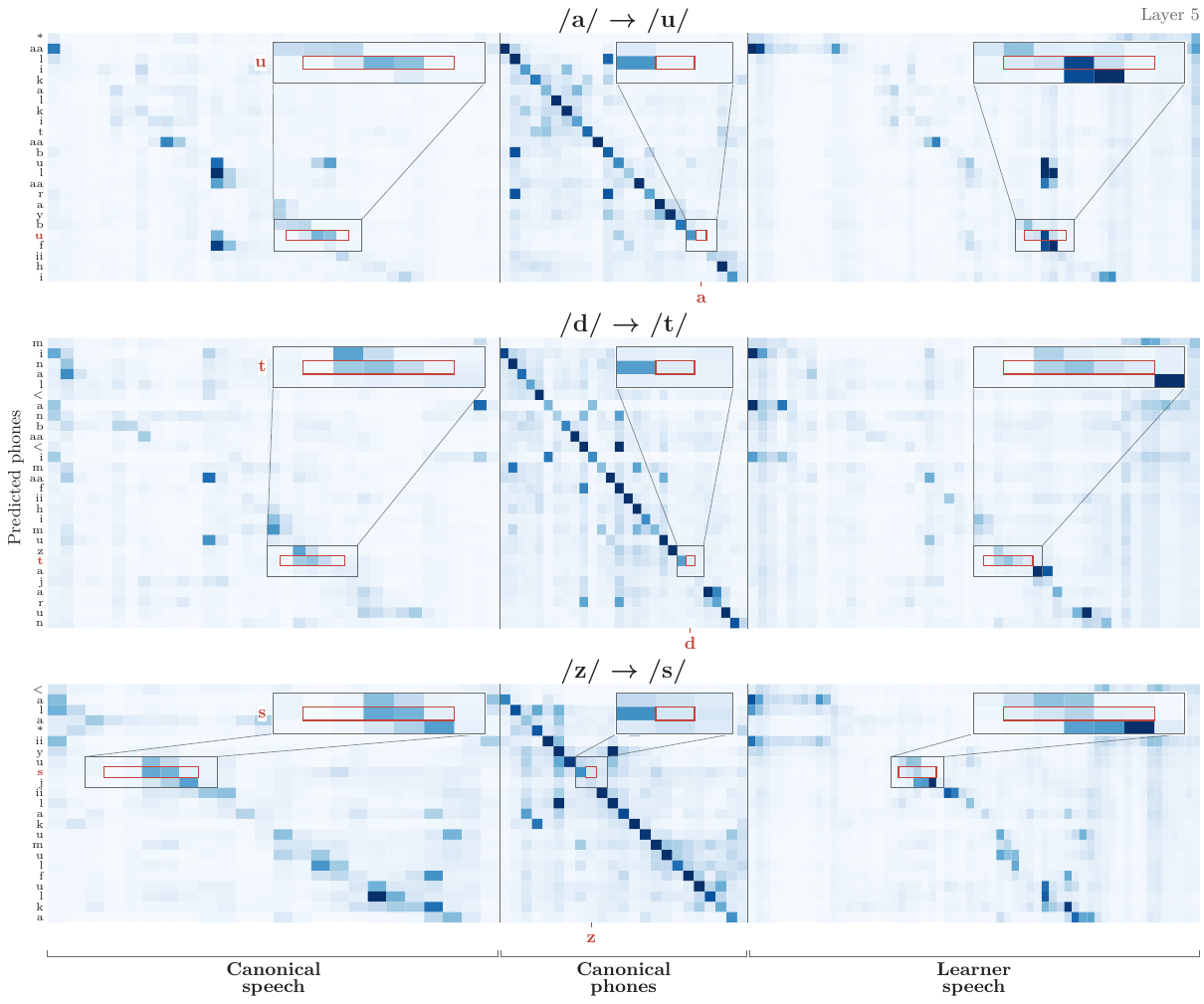}
\caption{
Layer-5 decoder attention of \textsc{X-MDD} for three Arabic
utterances. Insets magnify the gray-boxed regions, with red
boxes highlighting mispronounced output phones and red x-axis
labels marking their canonical counterparts.
Input segments are horizontally rescaled, with a separate
color scale for each utterance.
}
\label{fig:attention_maps}
\end{figure}

\paragraph{Attention patterns.}
Figure~\ref{fig:attention_share} shows that canonical phones
receive over half of the attention allocated to the three
inputs in layers 19--28.
The layer-5 maps in Figure~\ref{fig:attention_maps} reveal a
near-diagonal pattern, with attention broadly following
the canonical sequence as generation progresses.
Both speech inputs also receive localized attention,
including near the highlighted mispronunciations.
Together, these patterns suggest that
\textbf{canonical phones and canonical speech
jointly contribute to actual phone prediction}.

\begin{figure}[t]
\centering
\includegraphics[width=\columnwidth]{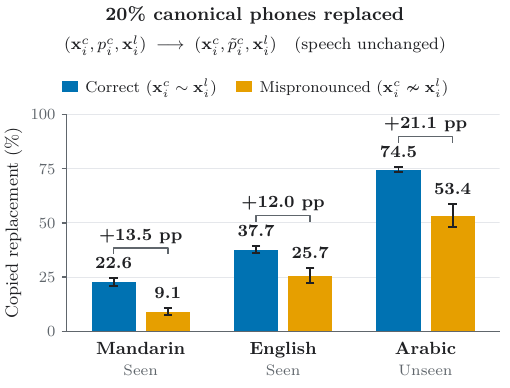}
\caption{Canonical phone intervention across three languages, showing how often the model follows the replaced phone at correctly pronounced and mispronounced positions. Error bars indicate 95\% confidence intervals.}
\label{fig:intervention}
\end{figure}

\paragraph{Canonical-phone corruption.}
We randomly replace 20\% of canonical phones while keeping
both speech inputs unchanged. Each replacement differs from
both the original canonical phone and the annotated actual
phone. For example, consider canonical phones
$\mathbf{p}^{c}=(\mathrm{K},\mathrm{AE},\mathrm{T})$ and actual phones
$\mathbf{p}^{a}=(\mathrm{D},\mathrm{AE},\mathrm{T})$.
Corrupting the originally correct second position yields
$\tilde{\mathbf{p}}^{c}=(\mathrm{K},\textcolor{green!75!black}{\mathrm{EH}},\mathrm{T})$,
whereas corrupting the originally mispronounced first position yields
$\tilde{\mathbf{p}}^{c}=(\textcolor{red!75!black}{\mathrm{P}},\mathrm{AE},\mathrm{T})$.
The canonical and learner speech agree at the former position
but differ at the latter; neither recording is modified.
We measure how often the model outputs the replacement phone,
grouping positions by their original pronunciation status.
As shown in Figure~\ref{fig:intervention},
\textbf{the model follows replacement phones more often at
originally correct positions} across all three languages.
This is consistent with a stronger canonical-phone prior
when canonical and actual pronunciation agree,
and reduced reliance on that prior at mispronounced positions. In other words, conflicting speech evidence reduces the tendency to copy canonical phones.  A qualitative example of this trend is provided in Case~3 of
Appendix~\ref{sec:case_studies}.

\subsection{Effects of GRPO on Phone Predictions}
\label{subsec:grpo_analysis}

\begin{figure}[!t]
\centering
\includegraphics[width=\columnwidth]{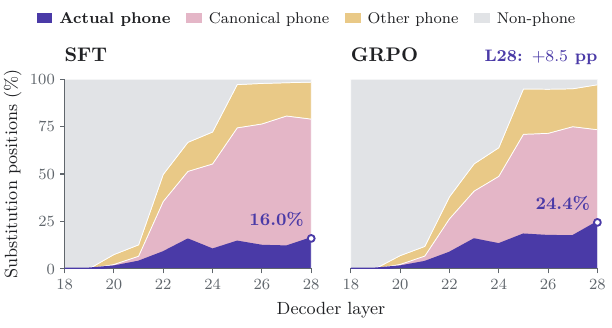}
\caption{
Logit-lens outcomes on Arabic substitution errors under
teacher forcing. At layer 28, the actual-phone share increases
from 16.0\% (SFT) to 24.4\% (GRPO), a gain of 8.5 percentage points.
}
\label{fig:lens_outcomes}
\end{figure}

\begin{figure}[!t]
\centering
\includegraphics[width=\columnwidth]{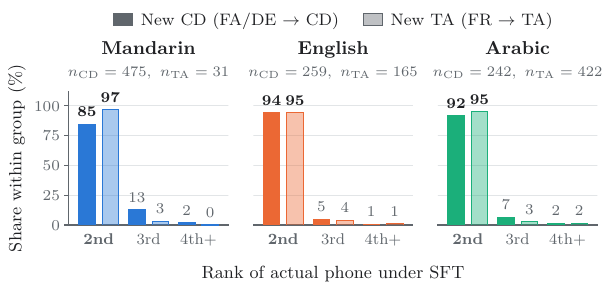}
\caption{
SFT ranks of actual phones newly corrected by GRPO under
teacher forcing. Solid/light bars denote new CD/TA, normalized
within each language and group. The 2nd bin includes rank-1
entries from the original grouping.
}
\label{fig:grpo_rank}
\end{figure}

Figure~\ref{fig:lens_outcomes} traces phone predictions through
decoder layers 18--28 using the logit lens on Arabic
substitution errors under teacher forcing. Here, substitution errors refer to positions where the annotated
actual phone differs from the canonical phone.
In later layers, GRPO increases actual phone predictions and
reduces canonical phone predictions, alongside a modest increase
in other phone predictions.
At layer 28, the actual phone share rises from 16.0\% under
SFT to 24.4\% after GRPO.
Figure~\ref{fig:grpo_rank} further examines the SFT ranks of
actual phones at positions newly corrected by GRPO.
Most were already ranked second by SFT, whereas lower-ranked phones account for few of these corrections.
Together, these results suggest that
\textbf{GRPO primarily promotes actual phones that are already highly ranked by SFT}. A qualitative example is provided in Case~1 of
Appendix~\ref{sec:case_studies}.

\section{Conclusion}
\label{sec:conclusion}

We presented \textsc{X-MDD}, a framework for transferring MDD
to languages unseen during MDD training using canonical speech
and phone references.
Trained on Mandarin and English, it transfers to Arabic with
canonical references and inference-time demonstrations.
Our analyses show that canonical phones and speech jointly
contribute to phone prediction.
The model relies more on canonical phones at correctly pronounced
positions, while conflicting speech evidence reduces this tendency.
GRPO further promotes actual phones already highly ranked by SFT.
Together, these results characterize how canonical references
and learner speech influence prediction.
Low recall and limited diagnosis accuracy on Arabic remain key
challenges.
Future work will evaluate
transfer across more unseen languages and learner populations.
\section{Limitations}

Evaluation on unseen languages is limited to Arabic, primarily due to the limited availability of publicly available MDD datasets. Furthermore, some accent-specific test sets contain only one speaker, making it difficult to disentangle accent-related differences from individual speaker characteristics and draw reliable conclusions about performance across accents. Although our results demonstrate the potential for generalization to an unseen language, broader evaluation across languages, accents, and learner populations is needed to establish the robustness and generalizability of the proposed framework. Future work could address these limitations by developing and releasing multilingual MDD evaluation datasets with greater speaker diversity and more balanced accent coverage, particularly for low-resource languages. Such efforts could facilitate more comprehensive evaluation of multilingual MDD systems and support more inclusive and accessible language learning.

\bibliography{custom}

\appendix

\label{sec:appendix}

\section{Comparison with Related Work}
\begin{table*}[t]
\centering
\small
\setlength{\tabcolsep}{5pt}
\caption{
Comparison with representative MDD systems.
CP and CS denote canonical phones and canonical speech,
respectively.
$\checkmark$: demonstrated;
$\triangle$: supported but not evaluated;
$\times$: not supported.}
\label{tab:related_work}
\begin{tabular}{lccc}
\toprule
Method & CP Input & CS Input & Multilingual \\
\midrule
ERN~\citep{DBLP:conf/icassp/LeeCG16}
    & \checkmark & $\times$ & $\triangle$ \\
CNN-RNN-CTC~\citep{leung2019cnn}
    & $\times$ & $\times$ & $\triangle$ \\
w2v-CTC~\citep{peng21e_interspeech}
    & $\times$ & $\times$ & $\triangle$ \\
VietMDD~\citep{huu23_interspeech}
    & \checkmark & $\times$ & $\triangle$ \\
L1-MultiMDD~\citep{DBLP:conf/icassp/KheirC024}
    & \checkmark & $\times$ & \checkmark \\
Pitch-Aware RNN-T~\citep{wang24la_interspeech}
    & $\triangle$ & $\times$ & $\triangle$ \\
ATP-LLM~\citep{wu24l_interspeech}
    & \checkmark & $\triangle$ & $\triangle$ \\
PER-MDD~\citep{tu2026mispronunciation}
    & \checkmark & $\times$ & $\triangle$ \\
LLM-MDD~\citep{xie2026mispronunciation}
    & $\checkmark$ & $\triangle$ & $\triangle$ \\
\midrule
\textbf{X-MDD (ours)}
    & \checkmark & \checkmark & \checkmark \\
\bottomrule
\end{tabular}
\end{table*}

\label{sec:related_comparison}

Table~\ref{tab:related_work} compares \textsc{X-MDD} with
representative MDD systems.
Most existing systems, such as ERN~\citep{DBLP:conf/icassp/LeeCG16},
VietMDD~\citep{huu23_interspeech}, and
ATP-LLM~\citep{wu24l_interspeech}, use canonical phones as input.
Recognition-based CTC systems, such as
CNN-RNN-CTC~\citep{leung2019cnn} and
w2v-CTC~\citep{peng21e_interspeech}, instead directly predict
the phones pronounced from learner speech.
In contrast, canonical speech has not been explicitly studied
as an input for MDD.
Recent LLM-based systems, such as ATP-LLM~\citep{wu24l_interspeech}
and LLM-MDD~\citep{xie2026mispronunciation}, could be extended
to incorporate canonical speech, but this setting has not been
evaluated.

Most existing MDD systems, such as
VietMDD~\citep{huu23_interspeech},
Pitch-Aware RNN-T~\citep{wang24la_interspeech}, and
PER-MDD~\citep{tu2026mispronunciation}, are designed for
individual languages.
L1-MultiMDD~\citep{DBLP:conf/icassp/KheirC024} is a notable
exception that jointly models multiple languages within a
unified MDD system.
\textsc{X-MDD} uses both canonical phone and speech references
in a multilingual model and further evaluates transfer to a
language excluded from MDD training.

\section{Datasets}

\label{sec:datasets_app}

\begin{table*}[t]
\centering
\small
\caption{Dataset statistics by language and accent. Each cell reports \textbf{Train $|$ Test}; Error denotes the percentage of mispronounced phones in the test set. Accents marked with $^\dagger$ are unseen during training.}
\label{tab:data}
\begin{adjustbox}{max width=\textwidth}
\begin{tabular}{lrrrr}
\toprule
Accent & Utterances & Speakers & Hours & Error (\%) \\
 
\midrule
\multicolumn{5}{l}{\textbf{Panel A: Mandarin}} \\
\midrule
\textit{AISHELL-3}
& 60,768 $|$ 0 & 169 $|$ 0 & 60.54 $|$ 0.00 & -- \\
\textit{Sichuan}
& 4,564 $|$ 300 & 8 $|$ 2 & 5.16 $|$ 0.27 & 38.03 \\
\textit{Singaporean}
& 2,882 $|$ 150 & 3 $|$ 1 & 2.11 $|$ 0.09 & 8.83 \\
\textit{Russian}
& 560 $|$ 150 & 1 $|$ 1 & 0.87 $|$ 0.24 & 22.21 \\
\textit{Korean}$^{\dagger}$
& 0 $|$ 150 & 0 $|$ 1 & 0.00 $|$ 0.25 & 10.15 \\
\textit{Chadian}$^{\dagger}$
& 0 $|$ 150 & 0 $|$ 1 & 0.00 $|$ 0.31 & 25.18 \\
\textbf{Total}
& \textbf{68,774 $|$ 900} & \textbf{181 $|$ 6} & \textbf{68.68 $|$ 1.16} & \textbf{27.06} \\
\midrule
\multicolumn{5}{l}{\textbf{Panel B: English}} \\
\midrule
\textit{TIMIT}
& 6,300 $|$ 0 & 630 $|$ 0 & 5.38 $|$ 0.00 & -- \\
\textit{Mandarin (SpeechOcean762)}
& 4,998 $|$ 0 & 250 $|$ 0 & 5.57 $|$ 0.00 & -- \\
\textit{Mandarin (L2-Arctic)}
& 450 $|$ 150 & 3 $|$ 1 & 0.49 $|$ 0.15 & 15.98 \\
\textit{Arabic}
& 449 $|$ 150 & 3 $|$ 1 & 0.46 $|$ 0.14 & 9.08 \\
\textit{Korean}
& 450 $|$ 150 & 3 $|$ 1 & 0.44 $|$ 0.15 & 9.69 \\
\textit{Vietnamese}
& 450 $|$ 150 & 3 $|$ 1 & 0.47 $|$ 0.17 & 23.82 \\
\textit{Spanish} 
& 450 $|$ 150 & 3 $|$ 1 & 0.53 $|$ 0.13 & 16.10 \\
\textit{Hindi} 
& 450 $|$ 150 & 3 $|$ 1 & 0.39 $|$ 0.14 & 11.22 \\
\textbf{Total}
& \textbf{13,997 $|$ 900} & \textbf{898 $|$ 6} & \textbf{13.73 $|$ 0.88} & \textbf{14.35} \\
\midrule
\multicolumn{5}{l}{\textbf{Panel C: Arabic}} \\
\midrule
\textit{QuranMB.v2} 
& 0 $|$ 1,643 & 0 $|$ 18 & 0.00 $|$ 1.92 & 6.73 \\

\bottomrule
\end{tabular}
\end{adjustbox}
\end{table*}

Table~\ref{tab:data} summarizes the open-sourcee datasets used for Mandarin, English, and Arabic.
For Mandarin and English, we combine native and non-native speech for training. The Mandarin training data include native speech from AISHELL-3~\citep{shi2020aishell} and MagicHub-Sichuan~\footnote{MagicHub-Sichuan: https://magichub.com/datasets/sichuan-dialect-scripted-speech-corpus-daily-use-sentence/}, and non-native speech from MagicHub-Singapore~\footnote{https://magichub.com/datasets/singaporean-chinese-conversational-speech-corpus/}, and LATIC~\citep{mqtj-qh10-21}. The English training data include native speech from TIMIT~\citep{garofolo1993timit} and non-native speech from SpeechOcean762~\citep{zhang2021speechocean762} and L2-ARCTIC~\citep{DBLP:conf/interspeech/ZhaoSSLCLG18}. All test speakers are unseen during training. For Mandarin, the Korean and Chadian accents are also unseen during training, as each contains only one speaker in the dataset and is therefore reserved for testing. Arabic is used only for unseen-language evaluation, and no Arabic data are used for training. For validation, we randomly select 500 utterances from the Mandarin training data and 500 from the English training data. For the Mandarin learner datasets, we randomly sample 300 or 150 utterances as the test set to match the size of the English test sets, and use the remaining utterances from the same speakers as the support set for same-speaker demonstrations.
\section{Experimental Details}
\subsection{Implementation Details}
\label{sec:details_app}
\paragraph{Audio inputs.} All input audio is resampled to 16~kHz. We use the
Qwen3-ASR feature extractor to compute 128-dimensional log-Mel filterbanks with a 25~ms window and a 10~ms hop size. 

\paragraph{Text tokens.} For Mandarin, the input text is converted to tone-marked Pinyin using G2PW~\footnote{G2PW: https://github.com/GitYCC/g2pW} for polyphone disambiguation. Each syllable is then decomposed into an initial and a tone-marked final, with tones 1--5, where tone 5 denotes the neutral tone. Zero-initial syllables are assigned an explicit placeholder based on the first letter of the final (\texttt{aa}, \texttt{ee}, \texttt{ii}, \texttt{oo}, \texttt{uu}, or \texttt{vv}), so that every syllable is represented as an initial--final pair. For English, we use the stress-free 39-phone CMU phone set. For Arabic, we use the official IqraEval phone annotations. We then tokenize the phone sequences using the original Qwen3 BPE tokenizer. Following the Qwen3-ASR output format, the model generates the predicted phone sequence after the language tag and \texttt{<asr\_text>} token. 
An example of the model input and output is shown below.



\begin{tcolorbox}[
    colback=gray!3,
    colframe=gray!50,
    colbacktitle=gray!12,
    coltitle=black,
    boxrule=0.5pt,
    arc=1pt,
    left=6pt,
    right=6pt,
    top=5pt,
    bottom=5pt,
    title={\textsc{X-MDD} input and output format},
    fonttitle=\small\bfseries,
    fontupper=\small,
    fontlower=\small,
    before upper={\raggedright},
    before lower={\raggedright},
]
\textbf{Input}\par
\smallskip
\texttt{<|im\_start|>user}\par
\texttt{<|task|>language: English; accent: Arabic}\par
\texttt{task: Predict the learner's actual phone sequence.}\par

\smallskip
{\color{gray!70!black}
\textit{Optional demonstrations ($N$ in total)}}\par
\texttt{<|demo\_start|>}\par
\texttt{<|canonical\_speech|>}
{\color{gray!70!black}\texttt{[audio features]}}\par
\texttt{<|canonical\_phone|>}\par
{\color{gray!70!black}\texttt{[canonical phones]}}\par
\texttt{<|learner\_speech|>}
{\color{gray!70!black}\texttt{[audio features]}}\par
\texttt{<|actual\_phone|>}\par
{\color{gray!70!black}\texttt{[annotated actual phones]}}\par
\texttt{<|demo\_end|>}\par
{\color{gray!70!black}\(\cdots\)}\par

\smallskip
{\color{gray!70!black}\textit{Query}}\par
\texttt{<|canonical\_speech|>}
{\color{gray!70!black}\texttt{[audio features]}}\par
\texttt{<|canonical\_phone|>}\par
\texttt{W IH L W IY EH V ER F AO R G EH T IH T}\par
\texttt{<|learner\_speech|>}
{\color{gray!70!black}\texttt{[audio features]}}\par
\texttt{<|im\_end|>}\par
\texttt{<|im\_start|>assistant}

\tcblower

\textbf{Output target}\par
\smallskip
\texttt{language English<asr\_text>}\par
\texttt{W IH L W IY EH V AH F AO R G EH T EH T}\par
\texttt{<|im\_end|>}

\smallskip
{\footnotesize\color{gray!70!black}
Gray text denotes annotations or placeholders;
audio features represent projected speech features.
}
\end{tcolorbox}
 
\paragraph{KNN-based Demonstrations.}
Since the English and Arabic test sets do not provide additional
same-speaker utterances for constructing support sets, we retrieve
demonstrations from SpeechOcean762 and \texttt{Iqra\_Extra\_IS26},
respectively.
We extract an x-vector for each utterance using WavLM%
~\footnote{\url{https://huggingface.co/microsoft/wavlm-base-plus-sv}}.
Audio is resampled to 16\,kHz, converted to mono, and truncated
to the first 10\,s, and each embedding is $\ell_2$-normalized.
Each test speaker is represented by the $\ell_2$-normalized mean
of the embeddings of their test utterances.
Pool utterances are ranked by cosine similarity to the test speaker,
and those with an actual phone sequence identical to that of the
query are excluded.
For each retrieved speech sample, we use it as the learner speech and synthesize its corresponding canonical speech for the demonstration.

\paragraph{TTS-based Demonstrations.} We conduct experiments by replacing the canonical and learner speech in the demonstrations at inference time as shown in Table~\ref{tab:icl_synthetic}. We next describe the construction of the TTS-based demonstrations in detail. For Mandarin, English and Arabic, we construct a TTS dataset with the same number of
utterances as the training set. Each real utterance produces one synthetic counterpart in which 50\% of the
canonical phone positions are retained and the remaining positions are corrupted. The corruption type
is sampled from substitution, deletion and insertion according to their empirical frequencies in the
real training annotations. For substitutions, the replacement phone is sampled from the empirical
confusion distribution conditioned on the canonical phone, while inserted phones are sampled from the
observed insertion distribution. All distributions are estimated from the corresponding training split
only.

For Mandarin, initials and finals are grouped into syllable-level units, and the sampled phone errors
are rendered as Chinese characters using empirical character--syllable mappings. For English, words
without selected errors remain in their orthographic form, whereas affected words are represented by
CosyVoice3 phone tokens, with substitutions, deletions and insertions applied directly at the selected
phone positions. Both Mandarin and English examples are synthesized using Fun-CosyVoice3-0.5B~\footnote{
\url{https://huggingface.co/FunAudioLLM/Fun-CosyVoice3-0.5B-2512}}. For Arabic, the sampled phone errors are rendered as
diacritized Arabic text and synthesized using SILMA TTS. The modified text is accepted only when its
Arabic G2P output matches the intended corrupted phone sequence; otherwise, the error pattern is
resampled. For all three languages, synthesis is conditioned on a reference recording
from the selected speaker. This preserves the selected speaker's timbre
while realizing the pronunciation specified by the synthesis text. All synthesized speech is generated at 24\,kHz and resampled to 16\,kHz before being passed to
the AuT.

\subsection{Baseline Models}
\label{sec:baselines}
 
We compare \textsc{X-MDD} with four monolingual MDD systems and one multilingual MDD system, all trained on the same training data for a fair comparison:

\begin{itemize}
    \item \textbf{w2v-CTC}~\citep{peng21e_interspeech}
    formulates MDD as CTC-based phone recognition, where the recognized phone sequence is compared
    with the canonical sequence to detect pronunciation errors. We use the best-performing
    multilingual configuration with wav2vec~2.0 XLSR-53 and jointly fine-tune the encoder and CTC head.

    \item \textbf{Pitch-aware RNN-T}~\citep{wang24la_interspeech}
    incorporates F0 information into a HuBERT-based stateless RNN-T to improve tonal-phone recognition.
    We follow the best-performing configuration with a 40-ms F0 hop size and jointly fine-tune HuBERT
    with the RNN-T model.

    \item \textbf{ATP-LLM}~\citep{wu24l_interspeech}
    conditions LLaMA~2-7B on learner speech, canonical phones, and text to generate the actual phone
    sequence. Speech is encoded by a wav2vec~2.0 Base encoder and projected into the LLM embedding space,
    with an auxiliary CTC objective and LoRA adaptation. We use the best-performing configuration
    with LoRA rank $R=32$.

    \item \textbf{WHU-IASP}~\citep{yang2026fusion}
    uses wav2vec~2.0 XLS-R-300M in a two-stage CTC framework. It combines hypotheses from multiple
    checkpoints through a confusion network and language-model rescoring for MDD.

    \item \textbf{L1-MultiMDD}~\citep{DBLP:conf/icassp/KheirC024}
    jointly performs multilingual MDD using speech, canonical phones, and learner L1 information.
    It uses wav2vec~2.0 XLSR-53 as the acoustic encoder and predicts phones over a unified multilingual phone set.
    We follow the best-performing sequential training setup reported in the original work.

\end{itemize}

\label{sec:implementation} 
\subsection{Evaluation Metrics}
\label{sec:metrics_app}

Following previous work~\citep{qian2010discriminative,yan2021end}, we evaluate both mispronunciation detection and diagnosis.
As shown in Table~\ref{tab:confusion_1}, true acceptance (TA) indicates that the pronunciation is correct and the model successfully predicts it as correct,
while false rejection (FR) indicates that a correct pronunciation is incorrectly predicted as a mispronunciation.
False acceptance (FA) indicates that a mispronunciation is incorrectly predicted as correct.
True rejection (TR) indicates that a mispronunciation is correctly detected, and is further categorized as a correct diagnosis (CD) when the predicted phone matches the actual phone, or a diagnosis error (DE) when the predicted phone differs from the actual phone.\textbf{ Precision, recall, and F1} are computed as: 
\begin{equation}
\begin{gathered}
P = \frac{\mathrm{TR}}{\mathrm{FR}+\mathrm{TR}}, \qquad
R = \frac{\mathrm{TR}}{\mathrm{FA}+\mathrm{TR}}, \\
F1 = \frac{2PR}{P+R}.
\end{gathered}
\end{equation}
\textbf{Diagnostic error rate (DER)} is computed as: 
\begin{equation}
\mathrm{DER} = \frac{\mathrm{DE}}{\mathrm{CD} + \mathrm{DE}}.
\end{equation}
We additionally report \textbf{phone error rate (PER)} for actual phone recognition:
\begin{equation}
\mathrm{PER} =
\frac{S+D+I}{N},
\end{equation}
where $S$, $D$, and $I$ denote the numbers of phone substitutions, deletions, and insertions,
respectively, and $N$ is the number of phones in the reference actual-phone sequence.


\begin{table}[t]
\centering
\small
\setlength{\tabcolsep}{6pt}
\renewcommand{\arraystretch}{1.15}
\caption{MDD confusion matrix. True rejections (TR) comprise
correct diagnoses (CD) and diagnosis errors (DE).}
\label{tab:confusion_1}
\begin{tabular}{@{}lcc@{}}
\toprule
\multirow{2}{*}{\textbf{Reference}}
& \multicolumn{2}{c}{\textbf{Prediction}} \\
\cmidrule(lr){2-3}
& Correct & Incorrect \\
\midrule
Correct   & TA & FR \\
Incorrect & FA & $\mathrm{TR}=\mathrm{CD}+\mathrm{DE}$ \\
\bottomrule
\end{tabular}
\end{table}
\section{Supplementary Results}


\subsection{Additional Results on Seen Languages}
\label{sec:appendix_seen}

\begin{table*}[t]
\centering
\small
\setlength{\tabcolsep}{3pt}
\renewcommand{\arraystretch}{1.08}
\caption{
Full ablation results on Mandarin (ZH) and English (EN),
with $N$ test-time demonstrations.
All values are percentages; P/R denote detection precision/recall.
\textbf{Bold}/\underline{underline} mark the best/second-best
results within each group and language, including ties.
}
\label{tab:seen_full}
\begin{tabular*}{\textwidth}{
@{\extracolsep{\fill}}lc*{5}{r}@{\hspace{6pt}}*{5}{r}@{}
}
\toprule
\multirow{2}{*}{Setting}
& \multirow{2}{*}{$N$}
& \multicolumn{5}{c}{\textbf{Mandarin (ZH)}}
& \multicolumn{5}{c}{\textbf{English (EN)}} \\
\cmidrule(lr){3-7}\cmidrule(lr){8-12}
& & PER$\downarrow$ & P$\uparrow$ & R$\uparrow$
& F1$\uparrow$ & DER$\downarrow$
& PER$\downarrow$ & P$\uparrow$ & R$\uparrow$
& F1$\uparrow$ & DER$\downarrow$ \\
\midrule
\multicolumn{12}{@{}l}{\textit{No canonical reference (learner speech only)}} \\
Base & 0
& \textbf{17.75} & \underline{81.85} & 77.25 & 79.49 & \textbf{31.67}
& 15.64 & 51.08 & 59.31 & 54.89 & 25.39 \\
\multirow{2}{*}{+ ICL} & 2
& 17.88 & 81.78 & 78.54 & 80.12 & 32.53
& \underline{15.48} & \underline{51.50} & 59.03 & 55.01 & 25.04 \\
 & 4
& \underline{17.86} & \textbf{81.97} & 78.54 & \underline{80.21} & \underline{32.38}
& \textbf{15.39} & \textbf{51.60} & 59.11 & 55.10 & 24.93 \\
\addlinespace[1pt]
+ GRPO & 0
& 18.53 & 80.22 & 78.64 & 79.42 & 32.94
& 16.69 & 48.69 & 60.96 & 54.14 & 25.62 \\
\multirow{2}{*}{+ ICL + GRPO} & 2
& 18.49 & 79.66 & \textbf{80.76} & \underline{80.21} & 33.14
& 16.58 & 48.71 & \textbf{66.01} & \textbf{56.06} & \textbf{23.77} \\
 & 4
& 18.32 & 79.95 & \underline{80.75} & \textbf{80.35} & 32.98
& 16.60 & 48.67 & \underline{65.43} & \underline{55.82} & \underline{24.26} \\
\midrule
\multicolumn{12}{@{}l}{\textit{Canonical phones}} \\
Base & 0
& \textbf{17.95} & \textbf{85.48} & 66.28 & 74.67 & 30.65
& 12.02 & 67.17 & 44.25 & 53.35 & \underline{18.10} \\
\multirow{2}{*}{+ ICL} & 2
& 19.84 & \underline{85.33} & 49.88 & 62.96 & \underline{26.86}
& \textbf{11.89} & \underline{67.22} & 46.02 & 54.64 & 18.12 \\
 & 4
& 20.01 & 85.16 & 48.43 & 61.75 & \textbf{25.51}
& \underline{11.90} & \textbf{67.30} & 45.08 & 54.00 & \textbf{18.05} \\
\addlinespace[1pt]
+ GRPO & 0
& \underline{18.58} & 81.17 & \textbf{75.13} & \textbf{78.03} & 33.56
& 14.27 & 54.18 & \textbf{59.01} & 56.49 & 20.60 \\
\multirow{2}{*}{+ ICL + GRPO} & 2
& 18.79 & 81.65 & \underline{69.84} & \underline{75.29} & 31.96
& 13.02 & 60.06 & \underline{53.70} & \textbf{56.70} & 20.99 \\
 & 4
& 18.94 & 81.26 & 69.02 & 74.64 & 31.52
& 12.89 & 60.68 & 52.92 & \underline{56.53} & 20.15 \\
\midrule
\multicolumn{12}{@{}l}{\textit{Canonical speech}} \\
Base & 0
& \textbf{17.17} & 85.18 & 70.72 & 77.28 & \textbf{29.66}
& 14.71 & 53.54 & 55.07 & 54.29 & \textbf{21.62} \\
\multirow{2}{*}{+ ICL} & 2
& \underline{17.54} & \textbf{85.76} & 69.00 & 76.47 & \underline{30.72}
& \textbf{14.40} & \textbf{54.52} & 54.44 & 54.48 & \underline{22.24} \\
 & 4
& 17.63 & \underline{85.54} & 68.70 & 76.20 & 30.75
& \underline{14.51} & \underline{53.98} & 54.13 & 54.06 & 22.41 \\
\addlinespace[1pt]
+ GRPO & 0
& 17.99 & 82.40 & \textbf{74.52} & \textbf{78.26} & 32.18
& 16.36 & 49.45 & \textbf{61.38} & \underline{54.78} & 23.14 \\
\multirow{2}{*}{+ ICL + GRPO} & 2
& 18.12 & 82.80 & \underline{73.19} & \underline{77.70} & 32.19
& 15.86 & 50.47 & 59.74 & 54.72 & 23.65 \\
 & 4
& 18.12 & 82.98 & 72.33 & 77.29 & 31.61
& 15.80 & 50.78 & \underline{60.42} & \textbf{55.18} & 23.80 \\
\midrule
\multicolumn{12}{@{}l}{\textit{Canonical speech + phones}} \\
Base & 0
& 18.21 & 84.86 & 65.46 & 73.91 & 30.45
& \underline{11.99} & 66.92 & 45.92 & 54.47 & 17.83 \\
\multirow{2}{*}{+ ICL} & 2
& \textbf{17.64} & \textbf{85.37} & 67.36 & 75.31 & \textbf{29.48}
& \textbf{11.64} & \textbf{68.87} & 45.51 & 54.81 & \underline{16.88} \\
 & 4
& 17.68 & \underline{85.34} & 68.29 & 75.87 & \underline{30.24}
& \textbf{11.64} & \underline{68.58} & 45.56 & 54.75 & \textbf{16.86} \\
\addlinespace[1pt]
+ GRPO & 0
& 18.99 & 78.72 & \textbf{79.97} & \textbf{79.34} & 34.16
& 12.91 & 60.57 & 56.10 & 58.25 & 20.36 \\
\multirow{2}{*}{+ ICL + GRPO} & 2
& 17.79 & 82.24 & 75.41 & 78.68 & 31.55
& 12.87 & 60.04 & \underline{57.57} & \underline{58.78} & 19.71 \\
 & 4
& \underline{17.66} & 82.46 & \underline{75.82} & \underline{79.00} & 31.32
& 12.86 & 60.21 & \textbf{57.62} & \textbf{58.88} & 19.47 \\
\midrule
\multicolumn{12}{@{}l}{\textit{Canonical speech + text}} \\
Base & 0
& 17.69 & 85.00 & 69.30 & 76.35 & \underline{30.83}
& 14.06 & 55.75 & 52.34 & 53.99 & \textbf{21.39} \\
\multirow{2}{*}{+ ICL} & 2
& \underline{17.54} & \underline{85.38} & 70.44 & 77.19 & 30.95
& \underline{14.03} & \underline{55.99} & 52.11 & 53.98 & 22.16 \\
 & 4
& \textbf{17.39} & \textbf{85.70} & 70.49 & 77.35 & \textbf{30.74}
& \textbf{13.97} & \textbf{56.12} & 51.88 & 53.92 & 22.16 \\
\addlinespace[1pt]
+ GRPO & 0
& 17.74 & 82.69 & 75.35 & 78.85 & 31.62
& 15.20 & 52.10 & \textbf{62.32} & \textbf{56.75} & \underline{21.78} \\
\multirow{2}{*}{+ ICL + GRPO} & 2
& 17.92 & 82.54 & \textbf{76.12} & \textbf{79.20} & 32.32
& 15.59 & 51.07 & \underline{60.70} & 55.47 & 22.31 \\
 & 4
& 17.89 & 82.89 & \underline{75.68} & \underline{79.12} & 32.35
& 15.43 & 51.47 & 60.30 & \underline{55.53} & 22.46 \\
\bottomrule
\end{tabular*}
\end{table*}

\label{sec:seen_full}

In this section, we provide additional experimental results to complement the findings presented in Section~\ref{subsec:seen_results}. As shown in Table~\ref{tab:seen_full}, we report detection precision and recall, along with results for both two- and four-shot ICL on the seen languages. The precision--recall results provide a more detailed view of the
effects of canonical references.
Canonical phones generally increase precision but reduce recall
before GRPO, indicating a stronger tendency to favor the canonical
pronunciation.
This effect is particularly pronounced on ZH, where adding four-shot
ICL to canonical phones increases precision from 85.48 to 85.16
while reducing recall from 66.28 to 48.43.
GRPO substantially recovers recall across canonical-reference
settings, often at the cost of lower precision.
For example, with canonical speech and phones, GRPO increases recall
from 65.46 to 79.97 on ZH and from 45.92 to 56.10 on EN.
These results further show that the F1 improvements from GRPO are
primarily associated with improved detection recall.

Increasing the number of demonstrations from two to four has a
relatively small and inconsistent effect.
For most settings, two- and four-shot ICL yield similar F1 scores,
and neither consistently dominates across ZH and EN.
For example, with canonical speech and phones plus GRPO, F1 changes
from 78.68 to 79.00 on ZH and from 58.78 to 58.88 on EN when
increasing $N$ from 2 to 4.
This suggests that most of the benefit of test-time demonstrations
is already obtained with a small number of examples, while additional
demonstrations provide only marginal gains on the seen languages.

\subsection{Additional Results on Unseen Languages}
\label{sec:appendix_unseen}

\begin{table*}[!t]
\centering
\small
\setlength{\tabcolsep}{4pt}
\renewcommand{\arraystretch}{1.10}
\caption{
Full ablation results on unseen Arabic with $N$ test-time
demonstrations. All metrics are percentages; P/R denote
detection precision/recall.
\textbf{Bold}/\underline{underline} mark the best/second-best
results, excluding the Arabic-trained baseline ($^\ast$)
and rows with PER $>100\%$.
}
\label{tab:unseen_full}
\begin{tabular}{@{}llcrrrrr@{}}
\toprule
Canonical references & Setting & $N$
& PER$\downarrow$ & P$\uparrow$ & R$\uparrow$
& F1$\uparrow$ & DER$\downarrow$ \\
\midrule

\multicolumn{2}{@{}l}{IqraEval baseline$^\ast$} & --
& 16.15 & 30.39 & 77.07 & 44.14 & 38.80 \\
\midrule

\multirow{4}{*}{\shortstack[l]{None\\(learner speech only)}}
& Base & 0
& 606.10 & 6.82 & 94.33 & 12.72 & 92.51 \\
& + ICL & 128
& 143.33 & 9.89 & 77.21 & 17.54 & 62.77 \\
& + GRPO & 0
& 371.74 & 7.27 & 88.91 & 13.44 & 87.29 \\
& + ICL + GRPO & 128
& 84.64 & 10.76 & \underline{74.29} & 18.80 & 58.93 \\
\midrule

\multirow{4}{*}{Phones}
& Base & 0
& 10.47 & 20.84 & 14.94 & 17.40 & 46.11 \\
& + ICL & 128
& 10.19 & 23.73 & 15.93 & 19.06 & \textbf{39.51} \\
& + GRPO & 0
& 9.94 & 28.50 & 27.05 & 27.75 & 43.88 \\
& + ICL + GRPO & 128
& 10.47 & 27.50 & 23.02 & 25.06 & 41.58 \\
\midrule

\multirow{4}{*}{Speech}
& Base & 0
& 462.54 & 6.73 & 92.22 & 12.55 & 93.08 \\
& + ICL & 128
& 156.84 & 9.35 & 76.38 & 16.66 & 70.95 \\
& + GRPO & 0
& 118.44 & 8.97 & 73.46 & 15.99 & 72.11 \\
& + ICL + GRPO & 128
& 94.38 & 10.10 & \textbf{75.54} & 17.82 & 66.80 \\
\midrule

\multirow{4}{*}{Speech + phones}
& Base & 0
& 10.15 & 34.54 & 18.25 & 23.88 & 50.98 \\
& + ICL & 128
& \textbf{7.63} & 40.72 & 17.15 & 24.13 & 46.26 \\
& + GRPO & 0
& 7.77 & \underline{48.24} & 29.76
& \underline{36.81} & 44.99 \\
& + ICL + GRPO & 128
& \underline{7.75} & \textbf{49.74} & 30.84
& \textbf{38.07} & \underline{40.33} \\
\midrule

\multirow{4}{*}{Speech + text}
& Base & 0
& 409.68 & 7.33 & 83.78 & 13.47 & 88.04 \\
& + ICL & 128
& 123.51 & 9.83 & 65.91 & 17.11 & 66.06 \\
& + GRPO & 0
& 110.81 & 9.52 & 70.03 & 16.76 & 72.70 \\
& + ICL + GRPO & 128
& 95.49 & 10.70 & 70.62 & 18.58 & 66.55 \\
\bottomrule
\end{tabular}
\end{table*}


\paragraph{Ablation study.}
We further evaluate MDD performance on unseen Arabic by reporting detection precision and recall and separately examining the effects of ICL and GRPO, as shown in Table~\ref{tab:unseen_full}.
The precision--recall results further clarify the different
roles of the canonical references.
Without canonical phones, the models exhibit very high recall but
extremely low precision and PER can become degenerate, indicating
substantial over-detection.
Canonical phones provide an important phonetic reference, greatly
improving precision and stabilizing phone prediction, although at
the cost of lower recall.
Canonical speech alone does not resolve this problem, but becomes
beneficial when combined with canonical phones, yielding a better
precision--recall balance and stronger overall performance.
The separate ICL and GRPO results further show that, with both
canonical references, ICL mainly improves precision and PER while
slightly reducing recall, whereas GRPO substantially improves recall
as well as precision.
Their combination achieves the strongest detection performance.
Overall, these results suggest that canonical phones provide the
essential symbolic reference for unseen-language transfer, while
canonical speech provides additional acoustic evidence, with ICL
and GRPO further refining the resulting predictions.

\begin{figure*}[!t]
    \centering
    \includegraphics[width=\textwidth]{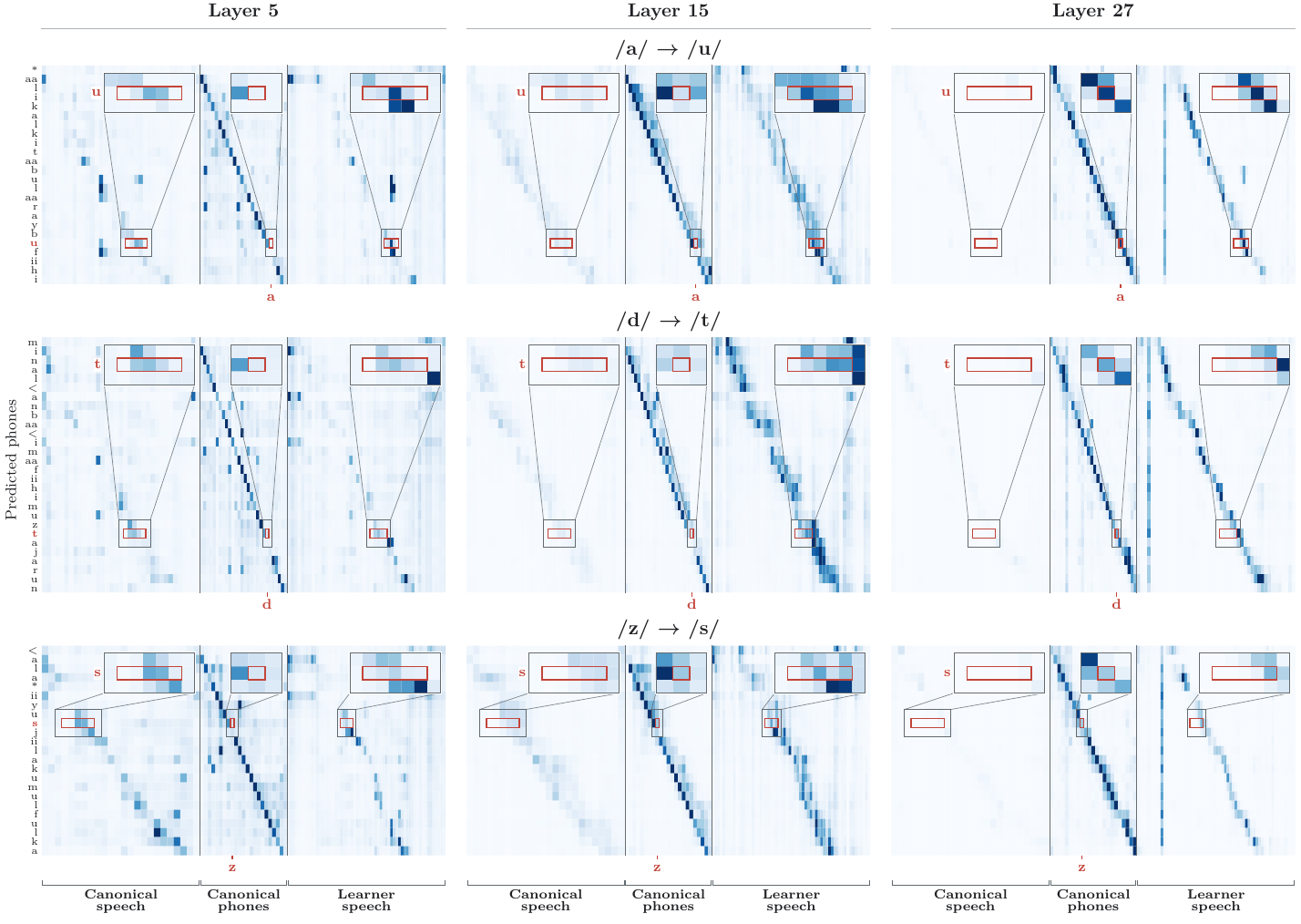}
    \caption{
    Decoder attention at layers 5, 15, and 27 (columns) for
    three Arabic utterances (rows). Insets enlarge matched
    regions across layers. Red boxes highlight mispronounced
    positions, and red ticks mark their canonical phones.
    Input segments are horizontally rescaled, with a separate
    color scale for each panel.
    }
    \label{fig:attention_layers}
\end{figure*}
 
\paragraph{Effect of constrained decoding.}
Table~\ref{tab:ar_constrained} examines whether the Arabic results
depend on restricting the decoder to the target phone inventory. OOV is the percentage of predicted phone symbols, pooled over all test utterances, that do not belong to the Arabic phone inventory (all phones in the QuranMB.v2 canonical and actual phone sequences).
Removing this constraint has only a minor effect on performance.
In particular, with canonical speech and phones, unconstrained
decoding achieves comparable PER and slightly higher F1, while only
0.12\% of the outputs are out-of-inventory symbols.
A similar pattern is observed with canonical phones alone, although
the OOV rate is slightly higher.
Constrained decoding slightly improves precision and DER, whereas
unconstrained decoding yields slightly higher recall.
Overall, the small differences between the two settings indicate that the unseen-language results are not primarily driven by the decoding constraint, as unconstrained decoding already produces almost exclusively valid target-language phones.
\begin{table}[t]
\centering
\footnotesize
\setlength{\tabcolsep}{2pt}
\renewcommand{\arraystretch}{1.08}
\caption{
Phone-constrained decoding on Arabic with ICL + GRPO
and $N=128$ demonstrations.
All values are percentages; OOV denotes out-of-inventory symbols.
Bold marks the better PER, F1, and DER within each group.
}
\label{tab:ar_constrained}
\begin{tabular*}{\columnwidth}{
@{\extracolsep{\fill}}lrrrrrr@{}
}
\toprule
Decoding
& PER$\downarrow$ & P$\uparrow$ & R$\uparrow$
& F1$\uparrow$ & DER$\downarrow$ & OOV$\downarrow$ \\
\midrule
\multicolumn{7}{@{}l}{\textit{Canonical phones}} \\
Constrained
& \textbf{10.47} & 27.50 & 23.02
& \textbf{25.06} & \textbf{41.58} & -- \\
Unconstrained
& 10.96 & 25.68 & 23.98
& 24.80 & 43.78 & 1.02 \\
\midrule
\multicolumn{7}{@{}l}{\textit{Canonical speech + phones}} \\
Constrained
& \textbf{7.75} & 49.74 & 30.84
& 38.07 & \textbf{40.33} & -- \\
Unconstrained
& 7.82 & 49.62 & 31.49
& \textbf{38.53} & 42.05 & 0.12 \\
\bottomrule
\end{tabular*}
\end{table}

\begin{table*}[!t]
\centering
\footnotesize
\setlength{\tabcolsep}{1.5pt}
\renewcommand{\arraystretch}{1.12}
\caption{
Demonstration-language ablation with inference-only ICL ($N=4$),
using demonstrations from other speakers.
Same/other refer to the query language; Other 1/2 denote
EN/AR for ZH, ZH/AR for EN, and ZH/EN for AR.
Mixed-language sampling is uniform subject to the stated counts.
All metrics are percentages; \textbf{bold}/\underline{underline}
mark the best/second-best PER, F1, and DER within each query language.
}
\label{tab:icl_multilingual}
\begin{tabular}{
@{}l*{5}{r}
@{\hspace{6pt}}*{5}{r}
@{\hspace{6pt}}*{5}{r}@{}
}
\toprule
\multirow{2}{*}{Demonstrations}
& \multicolumn{5}{c}{\textbf{Mandarin (ZH)}}
& \multicolumn{5}{c}{\textbf{English (EN)}}
& \multicolumn{5}{c}{\textbf{Arabic (AR)}} \\
\cmidrule(lr){2-6}
\cmidrule(lr){7-11}
\cmidrule(lr){12-16}
& PER$\downarrow$ & P$\uparrow$ & R$\uparrow$
& F1$\uparrow$ & DER$\downarrow$
& PER$\downarrow$ & P$\uparrow$ & R$\uparrow$
& F1$\uparrow$ & DER$\downarrow$
& PER$\downarrow$ & P$\uparrow$ & R$\uparrow$
& F1$\uparrow$ & DER$\downarrow$ \\
\midrule

Same language
& 17.56 & 82.36 & 76.29 & \textbf{79.21} & 31.23
& 12.97 & 59.81 & 57.62 & \textbf{58.69} & 20.13
& \textbf{8.61} & 37.55 & 30.81
& \textbf{33.85} & \underline{43.08} \\

3 same + 1 other
& 17.55 & 82.37 & 75.85 & 78.98 & \textbf{30.88}
& 13.01 & 59.67 & 55.67 & \underline{57.60} & 19.75
& \underline{8.82} & 36.53 & 31.17
& \underline{33.63} & 43.73 \\

2 same + 2 other
& 17.61 & 82.70 & 75.36 & 78.86 & 31.37
& 12.85 & 60.72 & 54.74 & 57.58 & 20.18
& 10.38 & 34.47 & 31.37 & 32.85 & 44.87 \\

1 same + 3 other
& \textbf{17.45} & 83.11 & 75.13 & 78.92 & 31.15
& 12.87 & 60.72 & 52.51 & 56.32 & 19.92
& 10.45 & 34.21 & 32.63 & 33.40 & \textbf{42.05} \\
\midrule

All-language mix
& 17.56 & 82.76 & 75.11 & 78.75 & 31.18
& 12.78 & 60.94 & 52.67 & 56.50 & \textbf{19.29}
& 10.67 & 33.70 & 31.70 & 32.67 & 44.59 \\

Other-language mix
& 17.47 & 83.17 & 74.97 & 78.86 & 31.20
& \underline{12.71} & 62.10 & 49.48
& 55.08 & \underline{19.66}
& 11.47 & 33.70 & 33.46 & 33.58 & 43.94 \\
\midrule

Other language 1
& 17.66 & 82.40 & 75.92 & \underline{79.03} & 31.72
& 13.03 & 61.11 & 46.93 & 53.09 & 21.92
& 10.83 & 34.17 & 30.15 & 32.03 & 45.90 \\

Other language 2
& \underline{17.46} & 83.99 & 73.02
& 78.12 & \underline{30.97}
& \textbf{12.48} & 63.71 & 47.16 & 54.20 & 20.15
& 11.97 & 30.09 & 34.27 & 32.04 & 43.34 \\

\bottomrule
\end{tabular}
\end{table*}
\subsection{Additional Analysis of Canonical Priors}
 \begin{figure*}[!t]
    \centering
    \includegraphics[width=1\linewidth]{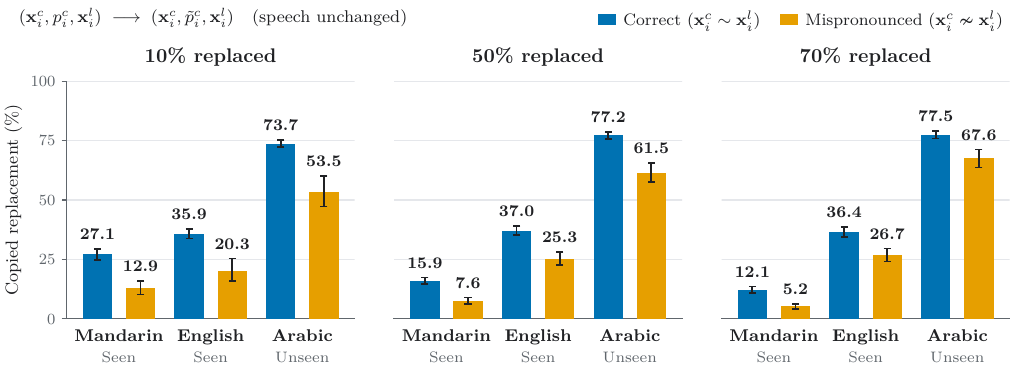}
    \caption{
Rates of copying replacement phones at intervention ratios
of 10\%, 50\%, and 70\%, grouped by original pronunciation
status. Both speech inputs remain unchanged.
Error bars indicate 95\% confidence intervals.
}
    \label{fig:intervention_ratio}
\end{figure*}
 
\paragraph{Layer-wise attention patterns.}
We provide additional attention visualizations for layers 15 and 27, as shown in Figure~\ref{fig:attention_layers}.
A clear layer-wise transition emerges across the three examples.
At layer 5, the decoder attends substantially to canonical speech,
with localized and near-diagonal patterns indicating alignment
between the generated phones and the canonical acoustic input.
As the decoder becomes deeper, attention to canonical speech
decreases, while canonical phones receive increasingly attention.
By layer 27, attention is concentrated primarily on canonical
phones, consistent with the aggregate trend in
Figure~\ref{fig:attention_share}.
Learner speech continues to receive localized attention across
layers, including around mispronounced positions.
These patterns suggest that canonical speech is primarily integrated
in earlier layers, while deeper layers increasingly rely on the
canonical phone sequence as a phonetic reference for actual phone
prediction.
Together with the attention-share analysis, this provides further
evidence that canonical speech and canonical phones contribute at
different stages of the decoder rather than through a single
shared attention pattern.

\paragraph{Canonical phone corruption ratios.}
In this section, we present additional results of the canonical phone intervention with replacement ratios of 10\%, 50\%, and 70\%, as shown in Figure~\ref{fig:intervention_ratio}.
Across all intervention ratios and all three languages, the model
consistently copies replacement phones more often at originally
correct positions than at mispronounced positions.
This pattern therefore remains stable across substantially different
levels of canonical phone corruption and is not specific to the
20\% setting used in the main analysis.
Although the absolute copying rates vary with the intervention ratio,
the gap between correctly pronounced and mispronounced positions
persists throughout.
These results further support the finding that the model relies more
strongly on canonical phones when canonical and learner speech agree,
whereas conflicting speech evidence at mispronounced positions
reduces this tendency.  More broadly, acoustic context has also been exploited to guide
speech reconstruction
\citep{zhao2026anybandunifiedmultibandwidthspeech}, highlighting the value of contextual acoustic
information.

\subsection{Additional Analysis of In-Context Learning}
\label{sec:appendix_icl}

 \begin{figure*}
     \centering
     \includegraphics[width=1\linewidth]{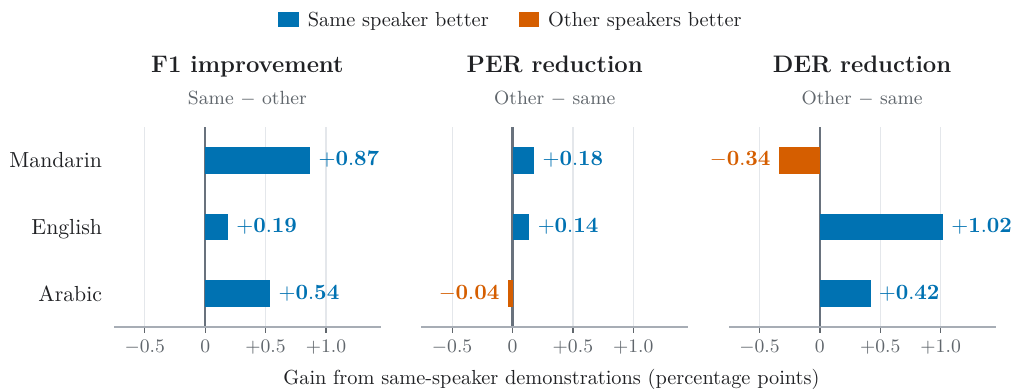}
     \caption{
Effect of same-speaker versus other-speaker demonstrations on
Mandarin, English, and Arabic.
We report $\Delta\mathrm{F1}=\mathrm{F1}_{\mathrm{same}}-\mathrm{F1}_{\mathrm{other}}$,
$\Delta\mathrm{PER}=\mathrm{PER}_{\mathrm{other}}-\mathrm{PER}_{\mathrm{same}}$,
and $\Delta\mathrm{DER}=\mathrm{DER}_{\mathrm{other}}-\mathrm{DER}_{\mathrm{same}}$,
so that positive $\Delta$ consistently indicates better performance
with same-speaker demonstrations.
All differences are in percentage points (pp).
}
     \label{fig:icl_speaker}
 \end{figure*}

\begin{table*}[!t]
\centering
\footnotesize
\setlength{\tabcolsep}{1.5pt}
\renewcommand{\arraystretch}{1.12}
\caption{
Real versus synthetic demonstrations for inference-only ICL
with \textsc{X-MDD} ($N=4$).
Real demonstrations come from the query speaker.
Synthetic demonstrations use synthesized canonical and learner
speech: CosyVoice error augmentation for ZH/EN and
\texttt{Iqra\_TTS} for AR.
All metrics are percentages; \textbf{bold}/\underline{underline}
mark the best/second-best PER, F1, and DER within each language,
including ties.
}
\label{tab:icl_synthetic}
\begin{tabular}{
@{}l*{5}{r}
@{\hspace{6pt}}*{5}{r}
@{\hspace{6pt}}*{5}{r}@{}
}
\toprule
\multirow{2}{*}{Demonstrations}
& \multicolumn{5}{c}{\textbf{Mandarin (ZH)}}
& \multicolumn{5}{c}{\textbf{English (EN)}}
& \multicolumn{5}{c}{\textbf{Arabic (AR)}} \\
\cmidrule(lr){2-6}
\cmidrule(lr){7-11}
\cmidrule(lr){12-16}
& PER$\downarrow$ & P$\uparrow$ & R$\uparrow$
& F1$\uparrow$ & DER$\downarrow$
& PER$\downarrow$ & P$\uparrow$ & R$\uparrow$
& F1$\uparrow$ & DER$\downarrow$
& PER$\downarrow$ & P$\uparrow$ & R$\uparrow$
& F1$\uparrow$ & DER$\downarrow$ \\
\midrule

4 real
& 17.72 & 82.36 & 75.40 & 78.72 & 31.18
& \underline{12.82} & 60.42 & 57.44
& \textbf{58.89} & 19.75
& \textbf{9.12} & 38.13 & 32.18 & 34.90 & 43.84 \\

3 real + 1 synthetic
& \textbf{17.57} & 82.57 & 75.55
& \textbf{78.90} & \textbf{30.91}
& \textbf{12.79} & 60.43 & 56.20
& \underline{58.24} & \textbf{19.20}
& \underline{9.27} & 37.76 & 32.96
& \textbf{35.20} & 43.62 \\

2 real + 2 synthetic
& \underline{17.66} & 82.34 & 75.46 & 78.75 & 31.17
& 12.87 & 60.38 & 54.49 & 57.28 & \underline{19.53}
& 10.08 & 35.81 & 33.82 & 34.79 & \underline{43.30} \\

1 real + 3 synthetic
& \textbf{17.57} & 82.69 & 75.28
& \underline{78.81} & \underline{31.02}
& 12.96 & 60.33 & 52.62 & 56.21 & 20.32
& 10.28 & 35.58 & 34.48 & 35.02 & 43.77 \\

4 synthetic
& 17.76 & 82.59 & 74.28 & 78.22 & 31.19
& 13.03 & 60.10 & 51.00 & 55.17 & 20.27
& 9.67 & 34.99 & 35.25
& \underline{35.12} & \textbf{43.15} \\

\bottomrule
\end{tabular}
\end{table*}
\paragraph{Effect of demonstration language.}
Table~\ref{tab:icl_multilingual} examines whether ICL depends on
the language of the test-time demonstrations.
Same-language demonstrations generally provide the strongest
detection performance, but the sensitivity to demonstration language
differs across query languages.
Mandarin is relatively insensitive to this choice, with similar F1
across same-language, mixed-language, and even other-language
demonstrations.
In contrast, English benefits more clearly from same-language
demonstrations, with F1 generally decreasing as they are replaced
by demonstrations from other languages.
Arabic shows a similar preference for same-language demonstrations,
although mixed- and other-language demonstrations remain competitive
in F1, indicating that useful contextual information can still
transfer across languages.
Overall, these results suggest that ICL is not purely
language-specific: cross-lingual demonstrations can provide useful
context, but same-language examples are generally more effective,
particularly for English and Arabic.

\paragraph{Effect of synthetic demonstrations.}
Table~\ref{tab:icl_synthetic} examines whether synthetic speech can
replace real test-time demonstrations.
The effect is strongly language-dependent.
For Mandarin, replacing real demonstrations with synthetic ones has
only a minor effect, with performance remaining stable across
different real--synthetic mixtures.
English is more sensitive to synthetic demonstrations: as the
proportion of synthetic examples increases, recall and F1 generally
decrease, while precision remains relatively stable.
In contrast, Arabic maintains comparable F1 even with fully synthetic
demonstrations.
Increasing the proportion of synthetic Arabic examples tends to
decrease precision but increase recall, resulting in a relatively
stable F1.
Overall, synthetic demonstrations can provide useful in-context
information and reduce the dependence on real speaker examples, but
their effectiveness varies across languages and synthesis settings.

\paragraph{Effect of speaker matching.}
Figure~\ref{fig:icl_speaker} compares same- and other-speaker
demonstrations sampled directly from the test sets, without
similarity-based retrieval. Same-speaker demonstrations are sampled from utterances that do not
overlap with the query utterance.
Same-speaker demonstrations generally perform better across the
three languages, as indicated by positive gains for most metrics.
They consistently improve F1, with the largest gain on Mandarin,
and also reduce PER for Mandarin and English.
DER is reduced for English and Arabic, while Mandarin is the only
case where same-speaker demonstrations slightly increase DER.
Overall, the predominance of improvements across languages and
metrics suggests that matching the demonstration speaker to the
query speaker provides useful information for ICL.
 
While same-speaker demonstrations generally improve
performance, it remains unclear whether these gains arise
from shared speaker timbre or pronunciation-related
characteristics. Future work could leverage disentangled
speech representations \citep{DBLP:conf/interspeech/ZhaoWW25,wang2026joycentmultiaccentttsdisentangled,zhao2026comelsinger}
to investigate the respective contributions of these
factors to speech-based ICL for MDD.
 
\section{Qualitative Case Studies}
\label{sec:case_studies}

\begin{figure*}
    \centering
    \includegraphics[width=1\linewidth]{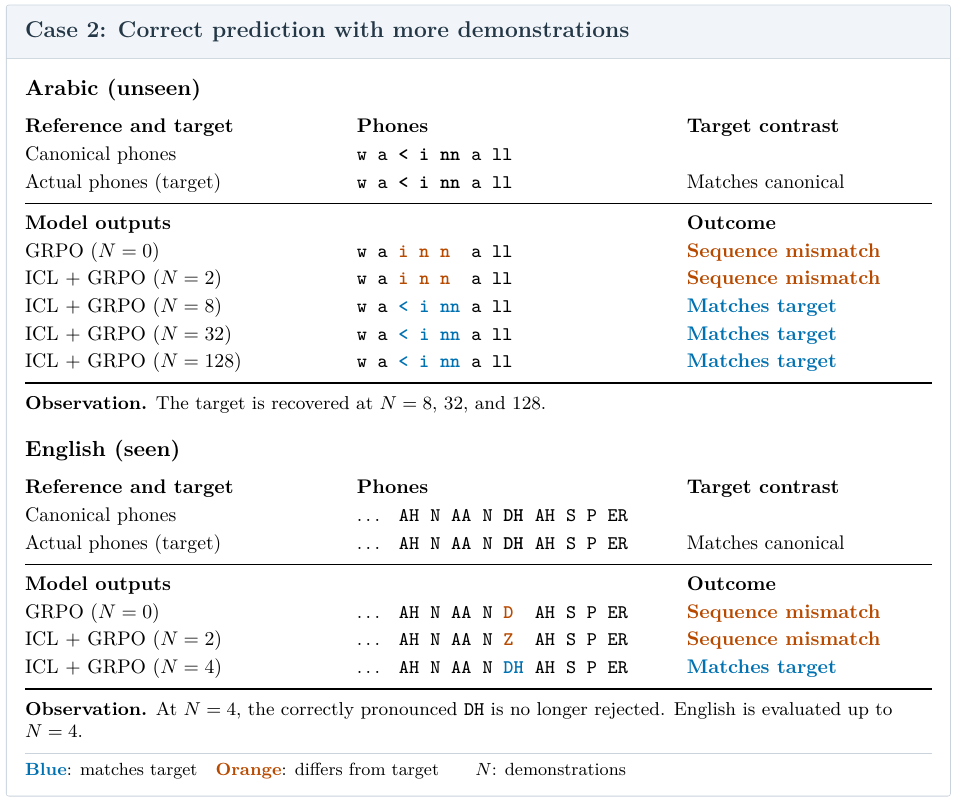}
    \caption{Qualitative examples of prediction errors corrected by increasing
the number of demonstrations on unseen Arabic and seen English.
With more demonstrations, the model recovers the annotated actual
phone sequence, which matches the canonical pronunciation in both
examples.}
    \label{fig:case2}
\end{figure*}
\begin{figure*}
    \centering
    \includegraphics[width=1\linewidth]{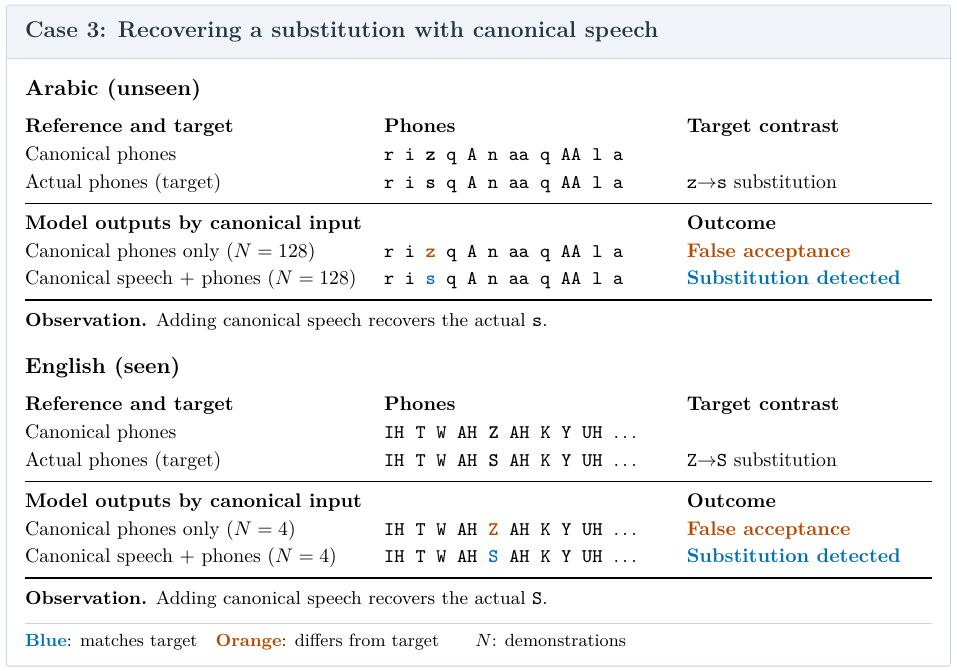}
    \caption{Qualitative examples of substitutions recovered by adding canonical
speech on unseen Arabic and seen English.
With canonical phones alone, the model falsely accepts the canonical
pronunciation, whereas adding canonical speech recovers the annotated
actual phone and correctly detects the substitution.}
    \label{fig:case3}
\end{figure*}

We furture provides qualitative examples of the behaviors
identified in our earlier analyses. In Figure~\ref{fig:case2} (Case~1), GRPO recovers false acceptances
by predicting the actual phone instead of copying the canonical phone.
Figure~\ref{fig:case2} (Case~2) illustrates the benefit of additional in-context demonstrations, where
the prediction becomes correct as more demonstrations are provided.
Figure~\ref{fig:case3} (Case~3) illustrates the complementary role of canonical speech: errors missed
with canonical phones alone are correctly recovered when canonical speech is
included. These examples provide concrete illustrations of the effects observed
in the quantitative analyses.

\begin{figure*}
    \centering
    \includegraphics[width=1\linewidth]{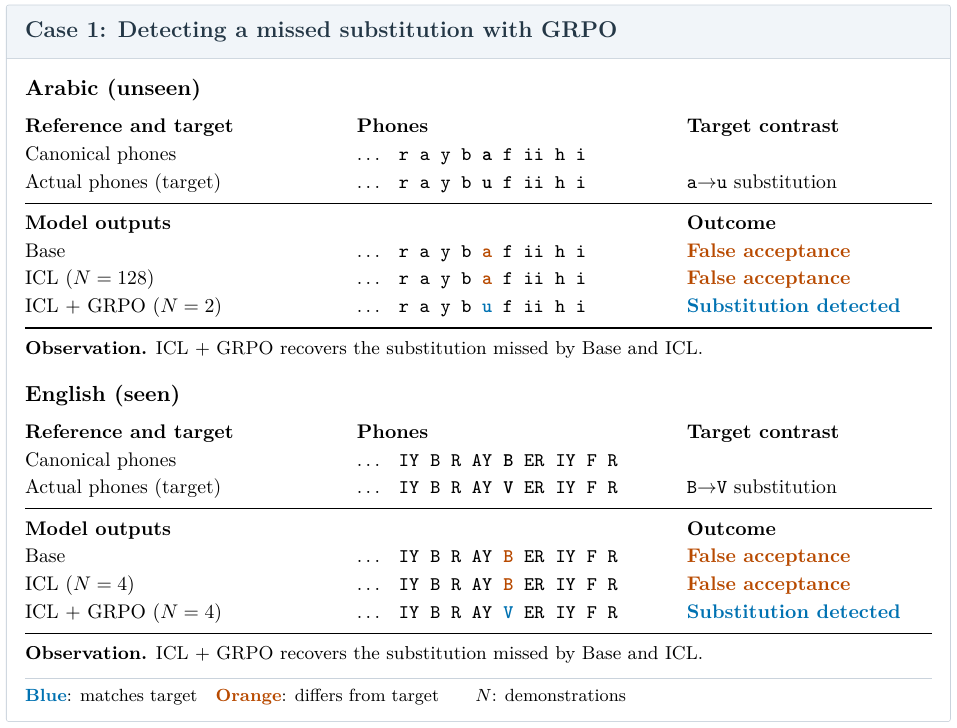}
    \caption{Qualitative examples of substitutions recovered by GRPO on
unseen Arabic and seen English.
Base and ICL falsely accept the canonical phone, whereas
ICL + GRPO predicts the annotated actual phone and correctly
detects the substitution.
Blue indicates phones matching the target, and orange indicates
phones differing from the target.}
    \label{fig:case1}
\end{figure*}

\section{Manual Annotation of Actual Phones}
\label{app:annotation}

We manually annotated three datasets, LATIC, MagicHub-Sichuan,
and MagicHub-Singapore, to obtain actual phone labels.
Three native Mandarin speakers were recruited as annotators.
To ensure annotation quality and consistency, we developed
automated scripts following
\citep{DBLP:conf/interspeech/ZhaoSSLCLG18}
to identify and correct annotation errors and resolve
ambiguities in the labels.
 
\end{document}